# $e^+e^- \to hhZ$ in the $B-L$ symmetric SSM

Dan-Dan Cui,[1,2,*] Tai-Fu Feng,[1,2,3,†] Yu-Li Yan,[1,2,‡] Hai-Bin Zhang,[1,2,§] Guo-Zhu Ning,[1,2] and Jin-Lei Yang[1,2,4,∥]

[1]*Department of Physics, Hebei University, Baoding 071002, China*
[2]*Key Laboratory of High-precision Computation and Application of Quantum Field Theory of Hebei Province, Baoding 071002, China*
[3]*Department of Physics, Chongqing University, Chongqing 400044, China*
[4]*CAS Key Laboratory of Theoretical Physics, School of Physical Sciences, University of Chinese Academy of Sciences, Beijing 100049, China*

The double Higgs boson production through $e^+e^- \to hhZ$ is analyzed in the minimal supersymmetric extension of the Standard Model (SM) with the local gauge symmetry $U(1)_{B-L}$, where $h$ denotes the lightest Higgs boson with 125 GeV. Considering the constraints from the updated prediction data, we find that the production cross section of this process in the model depends on some parameters strongly.

## I. INTRODUCTION

Among the particles predicted by the Standard Model (SM), the Higgs boson was the last particle discovered. Its discovery proves that correctness of the particles mass produced by the spontaneous symmetry breaking. Nevertheless, some other questions also arise. For example, is the discovered particle really the Higgs boson predicted by the SM? Are there other neutral or charged scalar fields? Whether or not the coupling of Higgs with matter fields and gauge particles meets the theoretical predictions of the SM? In addition, the SM can not provide the candidates for dark matter, can not explain the asymmetry of matter and antimatter in the universe, etc., so these require us to look for new physics beyond the SM. The $B-L$ Symmetric SM (B-LSSM) is one of the simplest extension models of the Minimal Supersymmetric Standard Model (MSSM), which is based on the gauge symmetry group $SU(3)_C \otimes SU(2)_L \otimes U(1)_Y \otimes U(1)_{B-L}$, where $B$ stands for the baryon number and $L$ stands for the lepton number, respectively.

The B-LSSM alleviates the hierarchy problem arisen in the MSSM, because the exotic singlet Higgs boson and right-handed (s)neutrinos [1–7] can alleviate the constraints from the experimental data of LHC, Tevatron and LEP. Furthermore, $U(1)_{B-L}$ gauge group can help to understand the possible broken ways of R parity in the supersymmetric models [8–10].

The cross section and decay branching ratios of the Higgs boson have been studied in various models [11–13]. Higgs pairs can be produced through gluon-gluon fusion in pp collision [14–16], WW or ZZ fusion [17–21], and so on. However the double Higgs-strahlung $e^+e^- \to hhZ$ is one of the main processes [22] for studying the self-coupling of Higgs boson.

In this work, we analyze the cross section of $e^+e^- \to hhZ$ in the B-LSSM. The cross section and angular distribution of this process can be used to determine the self-coupling of the Higgs at future collider experiments [23]. The process has been studied in the frameworks of the SM, the two-Higgs-doublet model [13,23–25], etc. In Ref. [13], the cross section in the SM was calculated as 0.16 fb at $E_{\mathrm{cm}} = 500$ GeV, 0.12 fb for $E_{\mathrm{cm}} = 1000$ GeV. If future experimental observations are much larger than the theoretical predictions of the SM, which can be considered as an evidence of new physics beyond the SM. Regarding the cross section, there are no experimental observations, but the production cross section for the $e^+e^- \to hhZ$ process is typically of the order of 0.1 fb at the collision energy just above the threshold at about 400 GeV, and at the international linear collider with a center-of-mass energy of 500 GeV, the trilinear Higgs boson coupling can be measured via this process [26]. In addition, if the experimental measurement value deviates obviously from the SM value in the future, the model can be considered as an explanation to account for the deviations. If the experimental measured values are consistent with the SM, it will constrain our parameter space.

*cuidandan0728@163.com
†fengtf@hbu.edu.cn
‡yychanghe@sina.com.cn
§hbzhang@hbu.edu.cn
∥yangjinlei@itp.ac.cn

The paper is organized as follows. In Sec. II, we introduce the B-LSSM in detail. In Sec. III, we analyze the dependence of the cross section and angular distribution of the final state particles on the parameters in this model. The numerical results are given in Sec. IV, and some conclusions are summarized in Sec. V.

## II. INTRODUCTION OF THE MODEL

In this section, we briefly introduce the basic characteristics of the B-LSSM. About the B-LSSM, there are several different versions. Here, we apply the version described in Refs. [27–30]. The B-LSSM of this version is encoded in SARAH [31–35], which can produce the mass matrices and interaction vertexes of this model. In this model, the local gauge group is enlarged to $SU(3)_C \otimes SU(2)_L \otimes U(1)_Y \otimes U(1)_{B-L}$, where the $U(1)_{B-L}$ is an additional gauge symmetry.

With the local gauge group $SU(3)_C \otimes SU(2)_L \otimes U(1)_Y \otimes U(1)_{B-L}$, the superpotential of the B-LSSM is written as

$$\mathcal{W} = \mathcal{W}_{\text{MSSM}} + \mathcal{W}_{(B-L)}. \tag{1}$$

Where $\mathcal{W}_{\text{MSSM}}$ denotes the superpotential of the MSSM [36], $\mathcal{W}_{(B-L)}$ denotes the additional terms in this model, and can be written as

$$\mathcal{W}_{(B-L)} = Y_{\nu,ij}\hat{L}_i\,\hat{H}_2\,\hat{\nu}^c_j + \mu'\hat{\eta}_1\,\hat{\eta}_2 + Y_{x,ij}\hat{\nu}^c_i\,\hat{\eta}_1\,\hat{\nu}^c_j. \tag{2}$$

The quantum numbers of the superfields of the quarks and leptons are assigned as

$$\begin{aligned}
&\hat{Q}_i = \begin{pmatrix} \hat{u}_i \\ \hat{d}_i \end{pmatrix} \sim (3,2,1/6,1/6),\\
&\hat{L}_i = \begin{pmatrix} \hat{\nu}_i \\ \hat{e}_i \end{pmatrix} \sim (1,2,-1/2,-1/2),\\
&\hat{U}^c_i \sim (3,1,-2/3,-1/6), \qquad \hat{D}^c_i \sim (3,1,1/3,-1/6),\\
&\hat{E}^c_i \sim (1,1,1,1/2),
\end{aligned} \tag{3}$$

with $i = 1,\ 2,\ 3$ denoting the generation indices. In addition, the quantum numbers of two Higgs doublets are

$$\begin{aligned}
\hat{H}_1 &= \begin{pmatrix} \hat{H}^1_1 \\ \hat{H}^2_1 \end{pmatrix} \sim (1,2,-1/2,0),\\
\hat{H}_2 &= \begin{pmatrix} \hat{H}^1_2 \\ \hat{H}^2_2 \end{pmatrix} \sim (1,2,1/2,0).
\end{aligned} \tag{4}$$

The quantum numbers of chiral singlet superfields are $\hat{\eta}_1 \sim (1,1,0,-1)$, $\hat{\eta}_2 \sim (1,1,0,1)$, and that of three generations of right-handed neutrinos is $\hat{\nu}^c_i \sim (1,1,0,1/2)$. Correspondingly, the soft breaking terms in the B-LSSM are written as

$$\mathcal{L}_{\text{soft}} = \mathcal{L}^{\text{MSSM}}_{\text{soft}} + \mathcal{L}^{B-L}_{\text{soft}}, \tag{5}$$

where the $\mathcal{L}^{\text{MSSM}}_{\text{soft}}$ denotes the soft breaking terms in the MSSM [36], and

$$\begin{aligned}
\mathcal{L}^{B-L}_{\text{soft}} = &\Big[-M_{BB'}\tilde{\lambda}_{B'}\tilde{\lambda}_B - \frac{1}{2}M_{B'}\tilde{\lambda}_{B'}\tilde{\lambda}_{B'} - B'\mu'\tilde{\eta}_1\tilde{\eta}_2\\
&+ T^{ij}_\nu H_2\tilde{\nu}^c_i\tilde{L}_j + T^{ij}_x\tilde{\eta}_1\tilde{\nu}^c_i\tilde{\nu}^c_j + \text{H.c.}\Big]\\
&- m^2_{\tilde{\nu},ij}(\tilde{\nu}^c_i)^*\tilde{\nu}^c_j - m^2_{\tilde{\eta}_1}|\tilde{\eta}_1|^2 - m^2_{\tilde{\eta}_2}|\tilde{\eta}_2|^2,
\end{aligned} \tag{6}$$

where $\lambda_B$, $\lambda'_B$ represent the gauginos of $U(1)_Y$ and $U(1)_{B-L}$, respectively. The local gauge symmetry $SU(3)_C\otimes SU(2)_L\otimes U(1)_Y\otimes U(1)_{B-L}$ is broken down to the electromagnetic symmetry $U(1)_{\text{em}}$ when the Higgs fields acquires the nonzero vacuum expectation values (VEVs):

$$\begin{aligned}
H^1_1 &= \frac{1}{\sqrt{2}}(v_1 + \text{ReH}^1_1 + i\text{Im}H^1_1),\\
H^2_2 &= \frac{1}{\sqrt{2}}(v_2 + \text{ReH}^2_2 + i\text{Im}H^2_2),\\
\tilde{\eta}_1 &= \frac{1}{\sqrt{2}}(u_1 + \text{Re}\tilde{\eta}_1 + i\text{Im}\tilde{\eta}_1),\\
\tilde{\eta}_2 &= \frac{1}{\sqrt{2}}(u_2 + \text{Re}\tilde{\eta}_2 + i\text{Im}\tilde{\eta}_2).
\end{aligned} \tag{7}$$

Similar to the ratio of nonzero VEVs of $H_1$ and $H_2$, we take $\tan\beta' = \frac{u_2}{u_1}$ denoting the ratio of nonzero VEVs of two chiral singlet superfields $\tilde{\eta}_1$ and $\tilde{\eta}_2$ here.

There is the gauge kinetic mixing $-\kappa_{Y,BL}A'^Y_\mu A'^{\mu,BL}$ from two local $U(1)$ gauge groups, and the mixing term satisfies the gauge invariance, where $A'^Y_\mu$, $A'^{\mu,BL}$ represent the gauge fields of two gauge groups $U(1)_Y$ and $U(1)_{B-L}$, respectively, and the antisymmetric tensor $-\kappa_{Y,BL}$ represents the mixing between two $U(1)$ gauge fields. The choice $\kappa_{Y,BL} = 0$ is unnatural because the mixing at low energy scale still can acquire a nonzero value through the evolution of renormalization group equations (RGEs) [37–43], even if we choose $\kappa_{Y,BL} = 0$ at the great uniform theory scale. The soft breaking parameters $\text{T}(T^{ij}_\nu, T^{ij}_x, T_{d,33}, T_{u,33})$ are proportional to the corresponding Yukawa couplings, i.e., $T^{ij}_\nu = Y_{\nu,ij}A_\nu$, $T^{ij}_x = Y_{x,ij}A_x$, $T_{d,33} = Y_bA_b$ and $T_{u,33} = Y_tA_t$ (the trilinear scalar terms in the soft supersymmetry breaking potential).

Because of the reasons above, the covariant derivative is usually written as

$$D_\mu = \partial_\mu - i(Y, B-L)\begin{pmatrix} g_Y & g'_{YB} \\ g'_{BY} & g_{B-L} \end{pmatrix} RR^T \begin{pmatrix} A'^Y_\mu \\ A'^{BL}_\mu \end{pmatrix}, \tag{8}$$

where $Y$, $B-L$ correspond to the hypercharge and $B-L$ charge, respectively. Furthermore, $R$ denotes a $2\times 2$

orthogonal matrix. We can choose a proper $R$ and then rewrite the coupling matrix as

$$\begin{pmatrix} g_Y & g'_{YB} \\ g'_{BY} & g_{B-L} \end{pmatrix} R = \begin{pmatrix} g_1 & g_{YB} \\ 0 & g_B \end{pmatrix}, \tag{9}$$

where $g_1$ is the hypercharge coupling constant of the SM, which can be modified in the B-LSSM. Meanwhile, two U(1) gauge fields are redefined as

$$\begin{pmatrix} A_\mu^Y \\ A_\mu^{BL} \end{pmatrix} = R^T \begin{pmatrix} A_\mu^{\prime Y} \\ A_\mu^{\prime BL} \end{pmatrix}. \tag{10}$$

The gauge kinetic mixing induces some interesting phenomenology. First, $A_\mu^{BL}$ boson can mix with the $A_\mu^Y$ and $V_\mu^3$ bosons at the tree level. In the interaction basis ($A_\mu^Y$, $V_\mu^3$, $A_\mu^{BL}$), the mass squared matrix of neutral gauge bosons is written as

$$\begin{pmatrix} \frac{1}{8}g_1^2v^2 & -\frac{1}{8}g_1g_2v^2 & \frac{1}{8}g_1g_{YB}v^2 \\ -\frac{1}{8}g_1g_2v^2 & \frac{1}{8}g_2^2v^2 & -\frac{1}{8}g_2g_{YB}v^2 \\ \frac{1}{8}g_1g_{YB}v^2 & -\frac{1}{8}g_2g_{YB}v^2 & \frac{1}{8}g_{YB}^2v^2 + \frac{1}{8}g_B^2u^2 \end{pmatrix}. \tag{11}$$

This mass squared matrix can be diagonalized by a unitary matrix, and the mass eigenstates can be written as linear combinations of ($A_\mu^Y$, $V_\mu^3$, $A_\mu^{BL}$):

$$\begin{pmatrix} \gamma_\mu \\ Z_\mu \\ Z'_\mu \end{pmatrix} = \begin{pmatrix} \cos\theta_W & \sin\theta_W & 0 \\ -\sin\theta_W\cos\theta'_W & \cos\theta_W\cos\theta'_W & \sin\theta'_W \\ \sin\theta_W\sin\theta'_W & -\cos\theta'_W\sin\theta'_W & \cos\theta'_W \end{pmatrix} \times \begin{pmatrix} A_\mu^Y \\ V_\mu^3 \\ A_\mu^{BL} \end{pmatrix}. \tag{12}$$

Here, $\theta_W$, $\theta'_W$ represent two mixing angles [44]:

$$\sin^2\theta_W = \frac{g_1^2}{g_1^2+g_2^2}, \tag{13}$$

$$\sin^2\theta'_W = \frac{1}{2} - \frac{4g_B^2[(g_{YB}^2-g_1^2-g_2^2)x^2+4g_B^2]}{[4g_{YB}^2(g_1^2+g_2^2)^2]x^4+8g_B^2(g_{YB}^2-g_1^2-g_2^2)x^2+32g_B^4}, \tag{14}$$

with $x=\frac{v}{u}$, $v^2=v_1^2+v_2^2$ and $u^2=u_1^2+u_2^2$. When $x \ll 1$, the eigenvalues of Eq. (11) can be written as [45]

$$\begin{aligned} &m_\gamma^2 = 0, \\ &m_Z^2 \simeq \frac{1}{4}(g_1^2+g_2^2)v^2 - \frac{1}{64g_B^2}(g_1^2+g_2^2+g_{YB}^2)^2x^2v^2, \\ &m_{Z'}^2 \simeq \frac{1}{8}\left[2g_{YB}^2v^2+8g_B^2u^2+\frac{1}{8g_B^2}(g_1^2+g_2^2+g_{YB}^2)^2x^2v^2\right]. \end{aligned} \tag{15}$$

The effective potential can be written as [46]:

$$V = V_0 + \Delta V_{1,t} + \Delta V_{2,\tilde{t}} + \Delta V_{3,\nu} + \Delta V_{4,\tilde{\nu}} + \Delta V_{5,b} + \Delta V_{6,\tilde{b}}. \tag{16}$$

Here, $V_0$ denotes the scalar potential at tree level, $\Delta V_{1,t}$ represents the correction from top quark, and $\Delta V_{2,\tilde{t}}$ represents the corrections from scalar top quarks, $\Delta V_{3,\nu}$ denotes the corrections from neutrinos, and $\Delta V_{4,\tilde{\nu}}$ denotes the corrections from sneutrinos $\Delta V_{5,b}$ represents the correction from bottom quark, and $\Delta V_{6,\tilde{b}}$ represents the corrections from scalar bottom quarks, respectively. The concrete expressions of those pieces are

$$\begin{aligned} V_0 = &\frac{1}{4}\left[\frac{1}{8}g^2(|H_1^1|^2-|H_2^2|^2)^2+\frac{1}{2}g_B^2(|\tilde{\eta}_1|^2-|\tilde{\eta}_2|^2)\right. \\ &\left.+\frac{1}{2}g_Bg_{YB}(|H_1^1|^2-|H_2^2|^2)(|\tilde{\eta}_1|^2-|\tilde{\eta}_2|^2)\right] \\ &+\frac{1}{2}[|\mu|^2(|H_1^1|^2+|H_2^2|^2)+|\mu'|^2(|\tilde{\eta}_1|^2+|\tilde{\eta}_2|^2) \\ &+m_{H_1}^2|H_1^1|^2+m_{H_2}^2|H_2^2|^2+m_{\tilde{\eta}_1}^2|\tilde{\eta}_1|^2+m_{\tilde{\eta}_2}^2|\tilde{\eta}_2|^2 \\ &+(-B'\mu'\tilde{\eta}_1\tilde{\eta}_2-B\mu H_1^1H_2^2+\text{H.c.})], \end{aligned} \tag{17}$$

$$\begin{aligned} \Delta V_{1,t} &= \frac{-3}{64\pi^2}m_t^4\left[\ln\left(\frac{m_t^2}{Q^2}\right)-\frac{3}{2}\right], \\ \Delta V_{2,\tilde{t}} &= \frac{3}{128\pi^2}\sum_{i=1}^2 m_{\tilde{t}_i}^4\left[\ln\left(\frac{m_{\tilde{t}_i}^2}{Q^2}\right)-\frac{3}{2}\right], \\ \Delta V_{3,\nu} &= \frac{-1}{64\pi^2}\sum_{i=1}^3 m_{\nu_{iR}}^4\left[\ln\left(\frac{m_{\nu_{iR}}^2}{Q^2}\right)-\frac{3}{2}\right], \\ \Delta V_{4,\tilde{\nu}} &= \frac{1}{256\pi^2}\sum_{i=1}^3 m_{\tilde{\nu}_{iR}}^4\left[\ln\left(\frac{m_{\tilde{\nu}_{iR}}^2}{Q^2}\right)-\frac{3}{2}\right], \\ \Delta V_{5,b} &= \frac{-3}{64\pi^2}m_b^4\left[\ln\left(\frac{m_b^2}{Q^2}\right)-\frac{3}{2}\right], \\ \Delta V_{6,\tilde{b}} &= \frac{3}{128\pi^2}\sum_{i=1}^2 m_{\tilde{b}_i}^4\left[\ln\left(\frac{m_{\tilde{b}_i}^2}{Q^2}\right)-\frac{3}{2}\right], \end{aligned} \tag{18}$$

where $Q$ denotes the renormalization scale; $m_t$, $m_b$ represent the masses of top and bottom quark, respectively,; and $m_{\tilde{t}_{1,2}}$, $m_{\tilde{b}_{1,2}}$ denote the masses of scalar top and scalar bottom quarks, respectively. In addition, $m_{\nu_{iR}}$ represents the masses of right-handed neutrinos, and $m_{\tilde{\nu}_{iR}}$ represents the masses of right-handed sneutrinos.

The stability conditions are

$$\frac{\partial V}{\partial \mathrm{ReH}_1^1}=0,\quad \frac{\partial V}{\partial \mathrm{ReH}_2^2}=0,\quad \frac{\partial V}{\partial \mathrm{Re}\tilde{\eta}_1}=0,\quad \frac{\partial V}{\partial \mathrm{Re}\tilde{\eta}_2}=0; \tag{19}$$

the detailed expression about Eq. (19) is given in Appendix A. Furthermore, the gauge kinetic mixing induces the mixing among the $H_1^1$, $H_2^2$, $\tilde{\eta}_1$, $\tilde{\eta}_2$ at tree level. In the interaction basis ($\mathrm{Re}H_1^1$, $\mathrm{Re}H_2^2$, $\mathrm{Re}\tilde{\eta}_1$, $\mathrm{Re}\tilde{\eta}_2$), the mass squared matrix for $CP$-even Higgs bosons is written as

$$M_h^2=\Delta m_h^2+\begin{pmatrix} \frac{1}{4}g^2v^2c_\beta^2+\mathrm{ReB}\mu t_\beta & -\frac{1}{4}g^2v^2s_\beta c_\beta-\mathrm{ReB}\mu & \frac{1}{2}g_Bg_{YB}vuc_\beta c_{\beta'} & -\frac{1}{2}g_Bg_{YB}vuc_\beta s_{\beta'} \\ -\frac{1}{4}g^2v^2s_\beta c_\beta-\mathrm{ReB}\mu & \frac{1}{4}g^2v^2s_\beta^2+\mathrm{ReB}\mu\, ctg\beta & \frac{1}{2}g_Bg_{YB}vuc_{\beta'}s_\beta & \frac{1}{2}g_Bg_{YB}vus_{\beta'}s_\beta \\ \frac{1}{2}g_Bg_{YB}vuc_\beta c_{\beta'} & \frac{1}{2}g_Bg_{YB}vuc_{\beta'}s_\beta & g_B^2u^2c_{\beta'}^2+\mathrm{ReB}'\mu' t_{\beta'} & -g_B^2u^2s_{\beta'}c_{\beta'}-\mathrm{ReB}'\mu' \\ -\frac{1}{2}g_Bg_{YB}vuc_\beta s_{\beta'} & \frac{1}{2}g_Bg_{YB}vus_{\beta'}s_\beta & -g_B^2u^2s_{\beta'}c_{\beta'}-\mathrm{ReB}'\mu' & g_B^2u^2s_{\beta'}^2+\mathrm{ReB}'\mu'\mathrm{ctg}\beta \end{pmatrix}, \tag{20}$$

where the abbreviations are $c_\beta=\cos\beta$, $s_\beta=\sin\beta$, $t_\beta=\tan\beta$, $c_{\beta'}=\cos\beta'$, $s_{\beta'}=\sin\beta'$, and

$$\Delta m_h^2=\begin{pmatrix} \Delta m_{\Phi_d\Phi_d} & \Delta m_{\Phi_u\Phi_d} & \Delta m_{\Phi_\eta\Phi_d} & \Delta m_{\Phi_{\bar\eta}\Phi_d} \\ \Delta m_{\Phi_d\Phi_u} & \Delta m_{\Phi_u\Phi_u} & \Delta m_{\Phi_\eta\Phi_u} & \Delta m_{\Phi_{\bar\eta}\Phi_u} \\ \Delta m_{\Phi_d\Phi_\eta} & \Delta m_{\Phi_u\Phi_\eta} & \Delta m_{\Phi_\eta\Phi_\eta} & \Delta m_{\Phi_{\bar\eta}\Phi_\eta} \\ \Delta m_{\Phi_d\Phi_{\bar\eta}} & \Delta m_{\Phi_u\Phi_{\bar\eta}} & \Delta m_{\Phi_\eta\Phi_{\bar\eta}} & \Delta m_{\Phi_{\bar\eta}\Phi_{\bar\eta}} \end{pmatrix}. \tag{21}$$

Here $\Delta m_h^2$ represents the one-loop correction to mass matrix squared; the detailed expression about this is given in Appendix A, and it can be obtained by the second derivative of the effective potential. In addition, $g^2=g_1^2+g_2^2+g_{YB}^2$. The mass matrix $M_h^2$ can be diagonalized by the $4\times4$ unitary matrix $Z^H$.

In the interaction basis ($\mathrm{ImH}_1^1,\mathrm{ImH}_2^2,\mathrm{Im}\tilde{\eta}_1,\mathrm{Im}\tilde{\eta}_2$), the mass squared matrix of $CP$-odd Higgs can be written as:

$$m_{A_0}^2=\begin{pmatrix} t_\beta B\mu & B\mu & 0 & 0 \\ B\mu & \frac{1}{t_\beta}B\mu & 0 & 0 \\ 0 & 0 & t_{\beta'}B'\mu' & B'\mu' \\ 0 & 0 & B'\mu' & \frac{1}{t_{\beta'}}B'\mu' \end{pmatrix}+\Delta m_{A_0}^2; \tag{22}$$

here, $\Delta m_{A_0}^2$ represents the one-loop correction to mass squared matrix of $CP$-odd Higgs, and $\Delta m_{A_0}^2=-\Delta m_h^2$. The mass matrix $m_{A_0}^2$ can be diagonalized by the $4\times4$ unitary matrix $Z^A$. The eigenvalues of Eq. (22) can be written as

$$\begin{aligned} m_{\eta^G_{1,2}}^2&=0,\\ m_{A_{1,2}}^2&=(m_{A_{1,2}}^{(0)})^2+\theta(\Delta m), \end{aligned} \tag{23}$$

where $m_{A_{1,2}}^{(0)}$ is the contribution under tree level approximation, and

$$\begin{aligned}\theta(\Delta m)=\frac{1}{2}[t\pm(t^2-4(&\Delta m_{\phi_d\phi_d}\Delta m_{\phi_u\phi_u}+\Delta m_{\phi_d\phi_d}\Delta m_{\phi_\eta\phi_\eta}+\Delta m_{\phi_u\phi_u}\Delta m_{\phi_\eta\phi_\eta}+(\Delta m_{\phi_d\phi_d}+\Delta m_{\phi_u\phi_u}+\Delta m_{\phi_\eta\phi_\eta})\Delta m_{\phi_{\bar\eta}\phi_{\bar\eta}}\\ &+\Delta m_{\phi_\eta\phi_d}^2+\Delta m_{\phi_{\bar\eta}\phi_u}^2))^{\frac{1}{2}}],\end{aligned} \tag{24}$$

with $t=\Delta m_{\phi_d\phi_d}+\Delta m_{\phi_u\phi_u}+\Delta m_{\phi_\eta\phi_\eta}+\Delta m_{\phi_{\bar\eta}\phi_{\bar\eta}}$; the detailed expression about $\theta(\Delta m)$ is given in Appendix A. The eigenstates corresponding to Goldstone are

$$\begin{aligned} A_{uj}^0&=c_\beta\mathrm{ImH}_1^1-s_\beta\mathrm{ImH}_2^2,\\ A_{\bar\eta j}^0&=c_{\beta'}\mathrm{Im}\tilde{\eta}_1-s_{\beta'}\mathrm{Im}\tilde{\eta}_2. \end{aligned} \tag{25}$$

The Higgs self-coupling can be defined as

$$
\begin{aligned}
\lambda_{h_ih_jh_k} = {} & \frac{1}{3!}\frac{\partial^3 V}{\partial(\mathrm{ReH}_1^1)^3} Z_{1i}^H Z_{1j}^H Z_{1k}^H + \frac{1}{3!}\frac{\partial^3 V}{\partial(\mathrm{ReH}_2^2)^3} Z_{2i}^H Z_{2j}^H Z_{2k}^H + \frac{1}{3!}\frac{\partial^3 V}{\partial(\mathrm{Re}\tilde{\eta}_1)^3} Z_{3i}^H Z_{3j}^H Z_{3k}^H + \frac{1}{3!}\frac{\partial^3 V}{\partial(\mathrm{Re}\tilde{\eta}_2)^3} Z_{4i}^H Z_{4j}^H Z_{4k}^H \\
& + \frac{1}{2!}\frac{\partial^3 V}{\partial(\mathrm{ReH}_1^1)^2\partial\mathrm{ReH}_2^2} Z_{1i}^H Z_{1j}^H Z_{2k}^H + \frac{1}{2!}\frac{\partial^3 V}{\partial(\mathrm{ReH}_1^1)^2\partial\mathrm{Re}\tilde{\eta}_1} Z_{1i}^H Z_{1j}^H Z_{3k}^H + \frac{1}{2!}\frac{\partial^3 V}{\partial(\mathrm{ReH}_1^1)^2\partial\mathrm{Re}\tilde{\eta}_2} Z_{1i}^H Z_{1j}^H Z_{4k}^H \\
& + \frac{1}{2!}\frac{\partial^3 V}{\partial(\mathrm{ReH}_2^2)^2\partial\mathrm{ReH}_1^1} Z_{2i}^H Z_{2j}^H Z_{1k}^H + \frac{1}{2!}\frac{\partial^3 V}{\partial(\mathrm{ReH}_2^2)^2\partial\mathrm{Re}\tilde{\eta}_1} Z_{2i}^H Z_{2j}^H Z_{3k}^H + \frac{1}{2!}\frac{\partial^3 V}{\partial(\mathrm{ReH}_2^2)^2\partial\mathrm{Re}\tilde{\eta}_2} Z_{2i}^H Z_{2j}^H Z_{4k}^H \\
& + \frac{1}{2!}\frac{\partial^3 V}{\partial(\mathrm{Re}\tilde{\eta}_1)^2\partial\mathrm{ReH}_1^1} Z_{3i}^H Z_{3j}^H Z_{1k}^H + \frac{1}{2!}\frac{\partial^3 V}{\partial(\mathrm{Re}\tilde{\eta}_1)^2\partial\mathrm{ReH}_2^2} Z_{3i}^H Z_{3j}^H Z_{2k}^H + \frac{1}{2!}\frac{\partial^3 V}{\partial(\mathrm{Re}\tilde{\eta}_1)^2\partial\mathrm{Re}\tilde{\eta}_2} Z_{3i}^H Z_{3j}^H Z_{4k}^H \\
& + \frac{1}{2!}\frac{\partial^3 V}{\partial(\mathrm{Re}\tilde{\eta}_2)^2\partial\mathrm{ReH}_1^1} Z_{4i}^H Z_{4j}^H Z_{1k}^H + \frac{1}{2!}\frac{\partial^3 V}{\partial(\mathrm{Re}\tilde{\eta}_2)^2\partial\mathrm{ReH}_2^2} Z_{4i}^H Z_{4j}^H Z_{2k}^H + \frac{1}{2!}\frac{\partial^3 V}{\partial(\mathrm{Re}\tilde{\eta}_2)^2\partial\mathrm{Re}\tilde{\eta}_1} Z_{4i}^H Z_{4j}^H Z_{3k}^H \\
& + \frac{\partial^3 V}{\partial\mathrm{ReH}_1^1\partial\mathrm{ReH}_2^2\partial\mathrm{Re}\tilde{\eta}_1} Z_{1i}^H Z_{2j}^H Z_{3k}^H + \frac{\partial^3 V}{\partial\mathrm{ReH}_1^1\partial\mathrm{ReH}_2^2\partial\mathrm{Re}\tilde{\eta}_2} Z_{1i}^H Z_{2j}^H Z_{4k}^H + \frac{\partial^3 V}{\partial\mathrm{ReH}_2^2\partial\mathrm{Re}\tilde{\eta}_1\partial\mathrm{Re}\tilde{\eta}_2} Z_{3i}^H Z_{2j}^H Z_{4k}^H \\
& + \frac{\partial^3 V}{\partial\mathrm{ReH}_1^1\partial\mathrm{Re}\tilde{\eta}_1\partial\mathrm{Re}\tilde{\eta}_2} Z_{3i}^H Z_{1j}^H Z_{4k}^H, \qquad (26)
\end{aligned}
$$

where $Y_t = \frac{\sqrt{2}m_t}{v_2}$, $Y_b = \frac{\sqrt{2}m_b}{v_1}$, $A_t$ is the trilinear couplings between Higgs and scalar top quarks, and $\mu$ denotes the mass parameter of Higgsino. In addition, the detailed expression about tree level correction $\lambda^{(0)}_{h_ih_jh_k}$ is given in Appendix B.

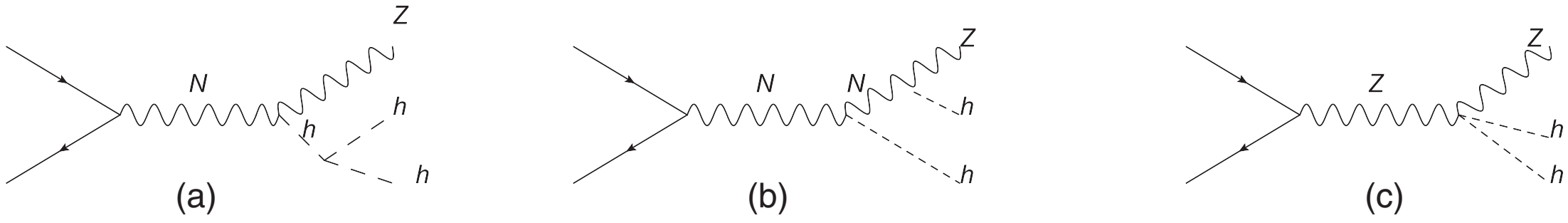

FIG. 1. Feynman diagrams for the Higgs boson pairs production through the process of $e^+e^- \to hhZ$ in the SM and in the B-LSSM.

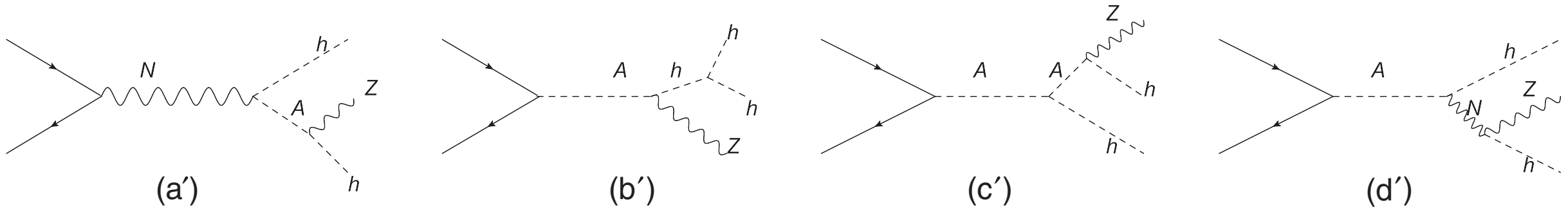

FIG. 2. Additional Feynman diagrams for the Higgs boson pairs production through the process of $e^+e^- \to hhZ$ in the B-LSSM.

The issues we discuss also involve the coupling of two $CP$-odd Higgs and one $CP$-even Higgs. The corresponding expression can be found in Appendix C.

## III. CROSS SECTION OF THE HIGGS BOSON PAIR PRODUCTION THROUGH $e^+e^- \to hhZ$

In this section, we will introduce the production of the Higgs boson pair through $e^+e^- \to hhZ$. The channel of the production of the Higgs boson pair is open when the collision energy $E_{\mathrm{cm}}$ of the initial state particle is more than about 340 GeV. The decay and production of Higgs boson have been discussed extensively [47–49]. In the framework of the B-LSSM, we mainly discuss the Higgs boson pair production through $e^+e^- \to hhZ$ here. We will carefully analyze the influence of relevant parameters on the total reaction cross section and angular distribution of the differential cross section in this model.

The Feynman diagrams contributing to this process are given in Figs. 1 and 2, where $N$ denotes $Z$ and $Z'$ bosons, $A$ represents $CP$-odd Higgs fields. The diagrams in Fig. 1 originate from the SM sector and the new physics sector, while those of Fig. 2 originate from the new physics sector, respectively.

In our calculation, we choose collision energy $E_{\mathrm{cm}} = 500$ GeV, so we ignore the masses of positron and electron. In addition, we neglect the Feynman diagrams generated by Yukawa couplings of electron.

In the B-LSSM, additional Feynman diagrams that contribute to this process are already given in Fig. 2. Taking Fig. 2($d'$) as an example, one derives the effective operator from the diagram ($d'$) as

$$\mathcal{O}_{4L,R} = \bar{v}(p_1) q_2^\alpha p_{L,R} u(p_2) \varepsilon_\alpha^*(k), \tag{27}$$

with $p_{L,R} = \frac{1 \mp \gamma^5}{2}$, where the $p_1$, $p_2$, $q_2$ represent the momenta of the initial state particles and final state Higgs boson, $k$ denotes the momentum of the final state vector boson, respectively. The corresponding effective amplitude can be written as:

$$M_{(2d')} = C_{4L}^{d'} \mathcal{O}_{4L} + C_{4R}^{d'} \mathcal{O}_{4R}. \tag{28}$$

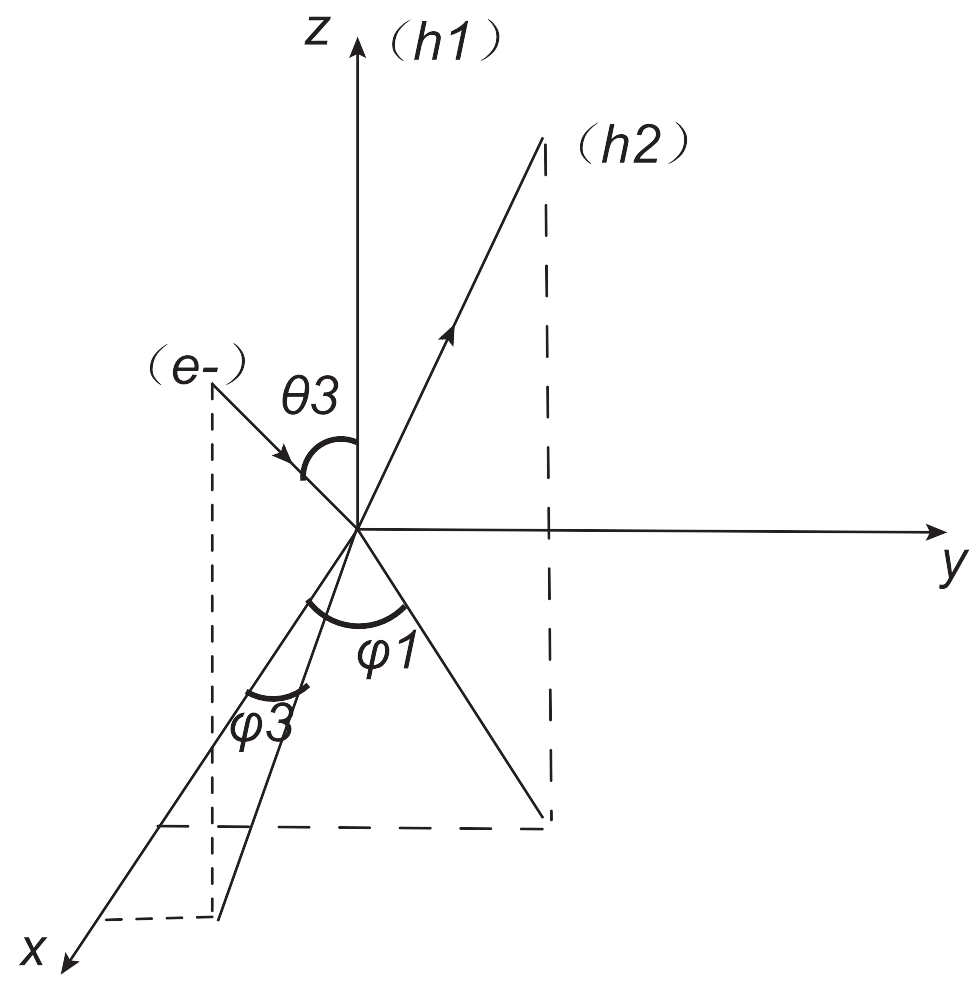

FIG. 3. The picture of kinematics.

The differential cross section of this process can be written as

$$\frac{d\sigma}{d\Omega} = \frac{1}{512\pi^5 E_{\rm cm}^2} \int_0^{2\pi} d\varphi_1 \int_{E_{1\,\rm min}}^{E_{1\,\rm max}} dE_1 \int_{E_{2\,\rm min}}^{E_{2\,\rm max}} dE_2 |M|^2. \tag{29}$$

Here, $M$ denotes the amplitude of all of these diagrams drawn in Figs. 1 and 2, and it is written as:

$$M = C_{L,R}^{(1)} \mathcal{O}_{L,R} + C_{L,R}^{(2)} \mathcal{O}_{L,R}; \tag{30}$$

the Wilson coefficients of those operators are given in Appendix D. In addition, $\varphi_1$ denotes the angle between the projection of the momentum direction of the final state Higgs on the x-y plane and the x axis, $\Omega$ is the spatial solid angle between the initial state electron and the final state Higgs, and $d\Omega = \sin\theta_3 d\theta_3 d\varphi_3$, respectively. We take the momentum direction of the final state Higgs as z axis, and the momentum direction of the initial state electron on the x-z plane. Here $\theta_3$ stands for the angle between initial state electron and final state Higgs, $\varphi_3$ is the angle between the projection of the momentum direction of the initial state electron on the x-y plane and the x axis, as shown in Fig. 3. Furthermore, $E_1$, $E_2$ both are the energy of the final state particles Higgs, where

$$\begin{aligned}
E_{1\,\rm min} &= m_h, \\
E_{1\,\rm max} &= \frac{E_{\rm cm}^2 - 2m_h m_z - m_z^2}{2E_{\rm cm}}, \\
E_{2\,\rm min} &= \frac{-(E_1 - E_{\rm cm})(2E_1 E_{\rm cm} - E_{\rm cm}^2 - 2m_h^2 + m_z^2)}{4E_1 E_{\rm cm} - 2(E_{\rm cm}^2 + m_h^2)} \\
&\quad + \frac{\sqrt{E_1^2 - m_h^2}\sqrt{(2E_1 E_{\rm cm} - E_{\rm cm}^2 + m_z^2)^2 - (2m_h m_z)^2}}{4E_1 E_{\rm cm} - 2(E_{\rm cm}^2 + m_h^2)}, \\
E_{2\,\rm max} &= \frac{-(E_1 - E_{\rm cm})(2E_1 E_{\rm cm} - E_{\rm cm}^2 - 2m_h^2 + m_z^2)}{4E_1 E_{\rm cm} - 2(E_{\rm cm}^2 + m_h^2)} \\
&\quad - \frac{\sqrt{E_1^2 - m_h^2}\sqrt{(2E_1 E_{\rm cm} - E_{\rm cm}^2 + m_z^2)^2 - (2m_h m_z)^2}}{4E_1 E_{\rm cm} - 2(E_{\rm cm}^2 + m_h^2)}.
\end{aligned} \tag{31}$$

In addition, the cross section about this process is

$$\sigma = \frac{1}{128\pi^4 E_{\rm cm}^2} \int_0^{2\pi} d\varphi_1 \int_{E_{1\,\rm min}}^{E_{1\,\rm max}} dE_1 \times \int_{E_{2\,\rm min}}^{E_{2\,\rm max}} dE_2 \int_0^1 |M|^2 d(\sin\theta_3). \tag{32}$$

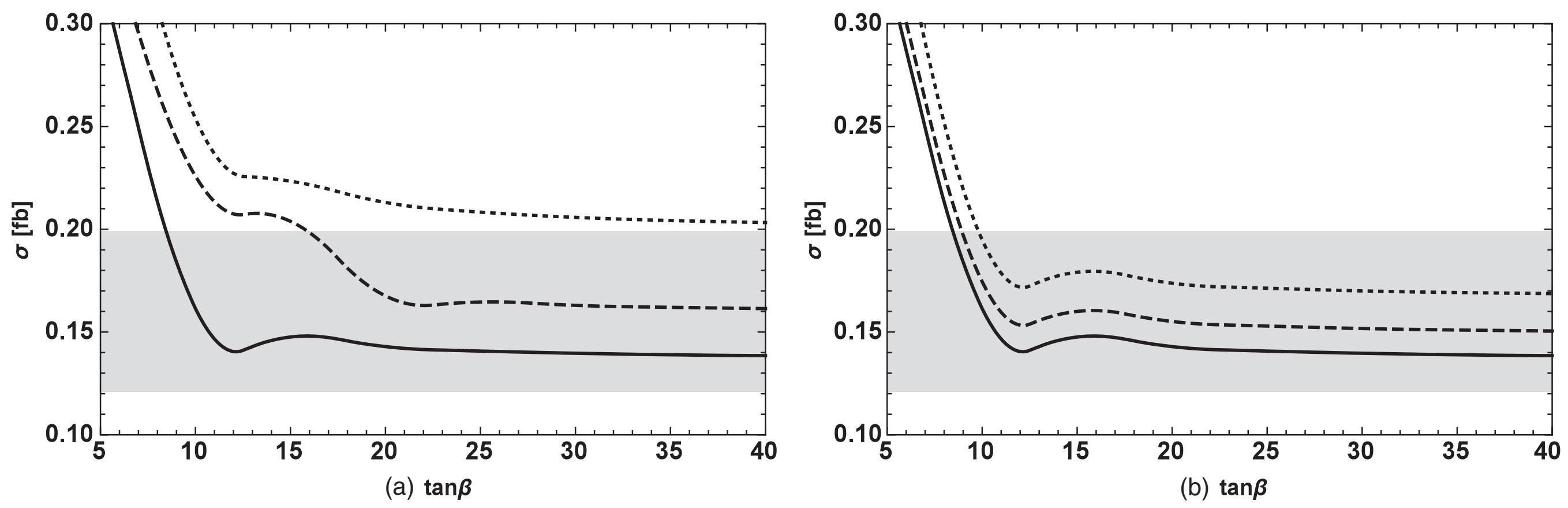

FIG. 4. The diagrams of cross section $\sigma$ versus $\tan\beta$, where (a) $g_{YB} = -0.9$ (solid line), $g_{YB} = -0.7$ (dashed line), $g_{YB} = -0.5$ (dotted line), and (b) $g_B = 0.7$ (solid line), $g_B = 0.5$ (dashed line), $g_B = 0.3$ (dotted line).

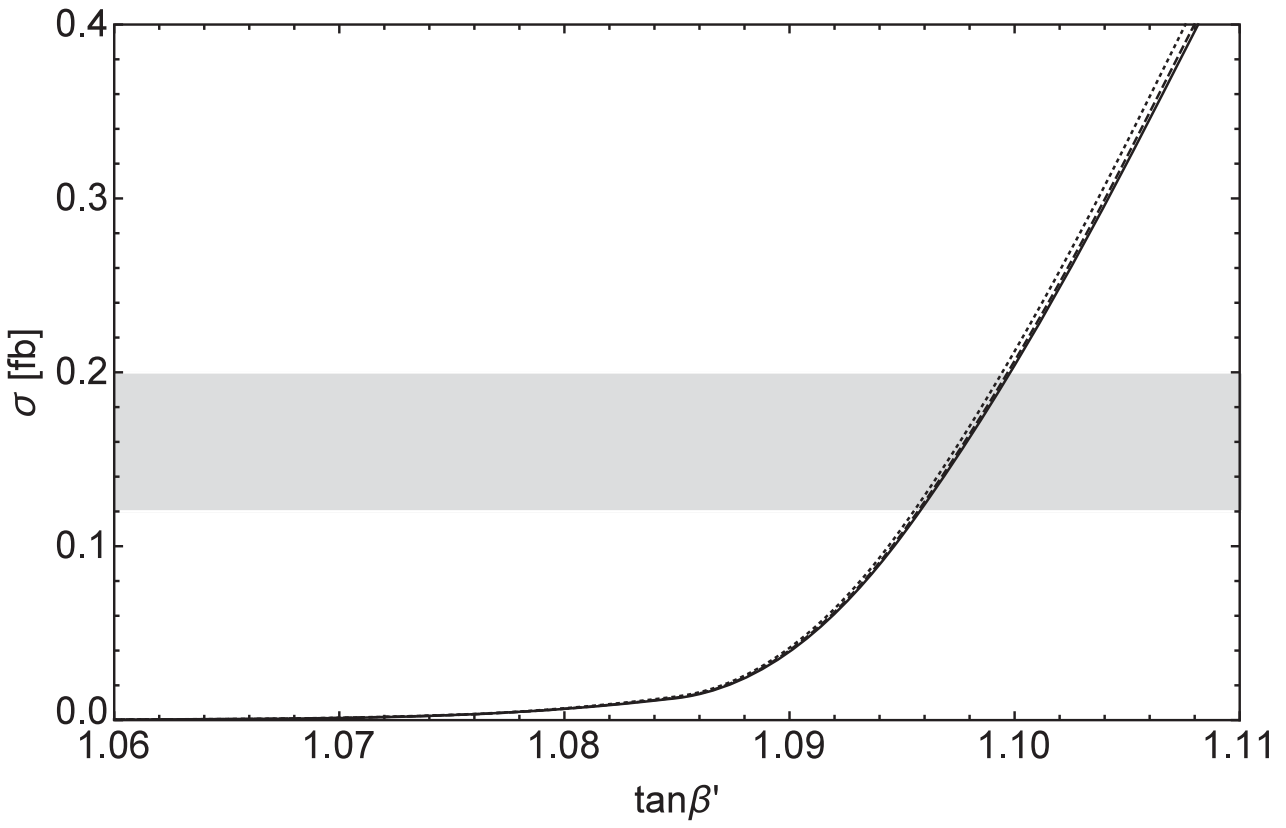

FIG. 5. The diagrams of cross section $\sigma$ versus $\tan\beta'$, where $g_B = 0.7$ (solid line), $g_B = 0.5$ (dashed line), $g_B = 0.3$ (dotted line).

## IV. NUMERICAL RESULTS

In this section, we will present the numerical results of the process $e^+e^- \to hhZ$. The input parameters related to the SM are chosen as $m_W = 80.385\,\text{GeV}$, $m_Z = 90.1876\,\text{GeV}$, $\alpha_{\text{em}}(m_Z) = 1/128.9$, $\alpha_s(m_Z) = 0.118$, $m_t = 173.5$ GeV, $m_b = 4.18$ GeV. The SM-like Higgs mass is [50]

$$m_h = 125.09 \pm 0.24 \text{ GeV}, \tag{33}$$

which constrains the parameter space of our model concretely [51]. We choose these parameters so that the corresponding theoretical prediction of the mass of the lightest $CP$-even Higgs fits the experimental data with 3 standard deviations: 124.37 GeV $\leq m_h \leq$ 125.81 GeV.

The updated experimental data [52] on searching $Z'$ indicates $M_{Z'} \geq 4.05$ TeV at 95% Confidence Level (CL), and we choose $M_{Z'} = 4.2$ TeV in our following numerical analysis. In addition, Refs. [53,54] give us a lower bound on the ratio between the $Z'$ mass and its gauge coupling at 99% CL as

$$M_{Z'}/g_B \geq 6 \text{ TeV}; \tag{34}$$

then the scope of $g_B$ is limited to $0 < g_B \leq 0.7$. The LHC experimental data also constrains the parameter $\tan\beta'$ as $\tan\beta' < 1.5$ [29]. In order to coincide with the constraints from the direct searches of the squarks at the LHC [55,56] and the observed Higgs signal in Ref. [57], for those parameters in the soft breaking terms, we take $A_\nu = A_x = 3$ TeV, $A_b = A_t = 1$ TeV, $\mu = 700$ GeV, $\mu' = 800$ GeV, $B\mu = 5\times10^5\,\text{GeV}^2$, $B'\mu' = 5\times10^5\,\text{GeV}^2$, $m_{\tilde{q}} = m_{\tilde{u}} = \text{diag}(2, 2, 1.6)$ TeV, and $u = \sqrt{u_1^2 + u_2^2}$; it can be obtained from Eq. (15). Now, we present our numerical results.

We adopt the latest predicted data of the SM to proceed in our analysis; it is about 0.12–0.20 fb [58]. Taking SM as a low energy effective theory, after integrating the heavy freedom of the high energy scalar, the new physical effect is constrained in some operators with dimensions greater than or equal to 6 composed of SM fields, where the operators with dimensions equal to 6 are given in Eq. (1) in the Ref. [58]. At present, the corresponding high-dimensional operator Wilson coefficients constrained by the electroweak precision test of the CMS and ATLAS experimental groups are as follows [58]:

$$\begin{aligned}
\bar{c}_T(m_Z) &\in [-1.5, 2.2] \times 10^{-3}, \\
(\bar{c}_W(m_Z) + \bar{c}_B(m_Z)) &\in [-1.4, 1.9] \times 10^{-3}, \\
\bar{c}_W \in [-0.05, 0.04], &\quad \bar{c}_{HW} \in [-0.1, 0.06], \\
\bar{c}_{HB} &\in [-0.05, 0.05].
\end{aligned} \tag{35}$$

Combining the tree diagrams and the correction of these high-dimensional operators, the SM theoretical prediction on $\sigma(e^+e^- \to hhZ) \simeq 0.160868 \pm 0.08$ fb can be obtained.

Taking $E_{\text{cm}} = 500$ GeV, $\tan\beta' = 1.1$, $m_{\tilde{\nu},33} = 600$ GeV, $m_{\tilde{L},33} = 600$ GeV, $Y_{\nu,33} = 5\times10^{-4}$, and $Y_{x,33} = 0.8$, we plot the total cross section $\sigma$ of $e^+e^- \to hhZ$ versus the parameter $\tan\beta$ in the Fig. 4, where the gray band represents the SM prediction with 3 standard deviations [58]. In the Fig. 4(a), we take $g_B = 0.7$, and $g_{YB} = -0.9$ (solid line), $g_{YB} = -0.7$ (dashed line), $g_{YB} = -0.5$ (dotted line), respectively. In the Fig. 4(b), we take $g_{YB} = -0.9$, and $g_B = 0.7$ (solid line), $g_B = 0.5$ (dashed line), $g_B = 0.3$ (dotted line), respectively. Obviously the theoretical prediction on the total cross section $\sigma$ depends on the parameter $\tan\beta$ strongly. Along with the increasing of $\tan\beta$, the cross section decreases steeply as $\tan\beta \leq 15$. As $\tan\beta > 20$, the dependence of total cross section on $\tan\beta$ is mild. Furthermore, the difference between the prediction from the SM and that from the B-LSSM exceeds 3 standard deviations.

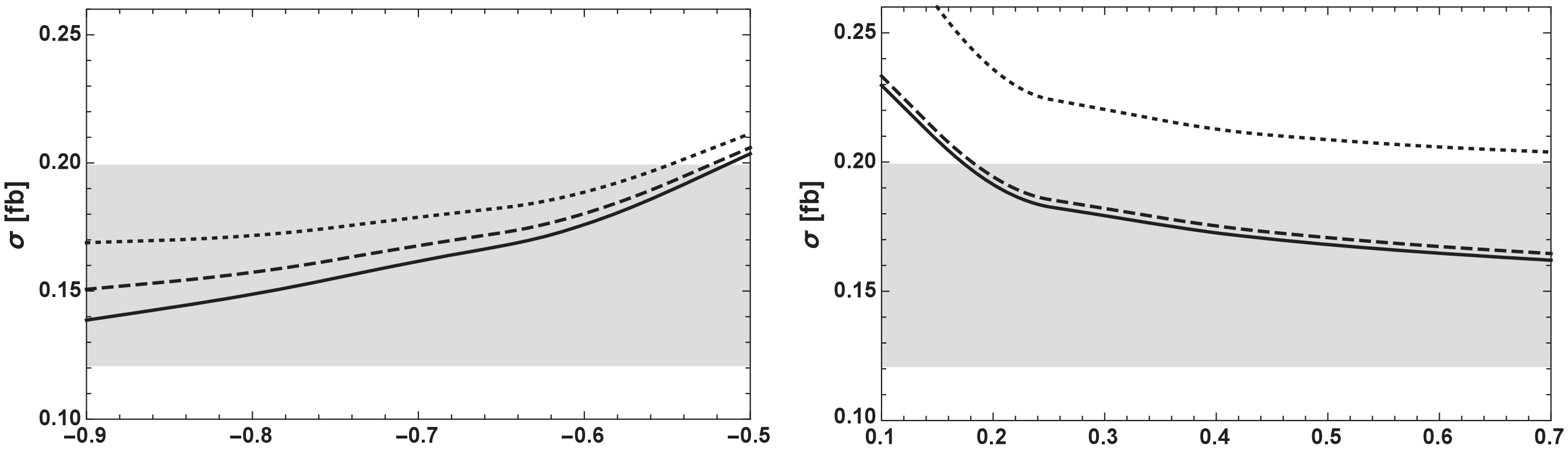

(a) $g_{YB}$ (b) $g_B$

FIG. 6. The diagrams of cross section $\sigma$ versus $g_B$ and $g_{YB}$, where (a) $g_B = 0.7$ (solid line), $g_B = 0.5$ (dashed line), $g_B = 0.3$ (dotted line), and (b) $\tan\beta = 35$ (solid line), $\tan\beta = 25$ (dashed line), $\tan\beta = 15$ (dotted line).

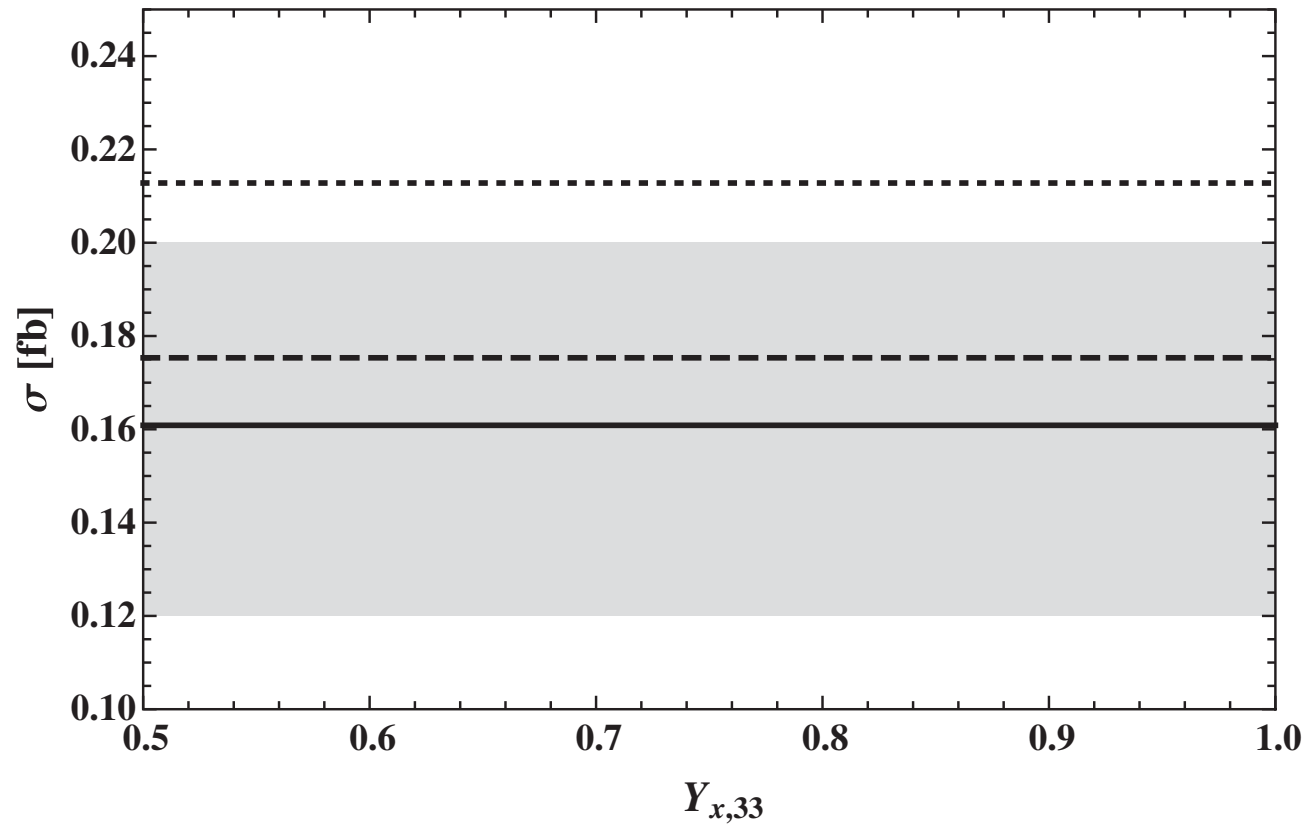

FIG. 7. The diagrams of cross section $\sigma$ versus $Y_{x,33}$, where $g_{YB} = -0.9$ (solid line), $g_{YB} = -0.7$ (dashed line), $g_{YB} = -0.5$ (dotted line).

Taking $E_{\rm cm} = 500$ GeV, $\tan\beta = 35$, $g_{YB} = -0.5$, $m_{\tilde{\nu},33} = 600$ GeV, $m_{\tilde{L},33} = 600$ GeV, $Y_{\nu,33} = 5 \times 10^{-4}$, and $Y_{x,33} = 0.8$, we plot the total cross section $\sigma$ of $e^+e^- \to hhZ$ versus the parameter $\tan\beta'$ in the Fig. 5, where the gray band represents the SM prediction with 3 standard deviations [58]. In the Fig. 5, $g_B = 0.7$ (solid line), $g_B = 0.5$ (dashed line) and $g_B = 0.3$ (dotted line), respectively. With the increasing of $\tan\beta'$, the cross section increases steeply.

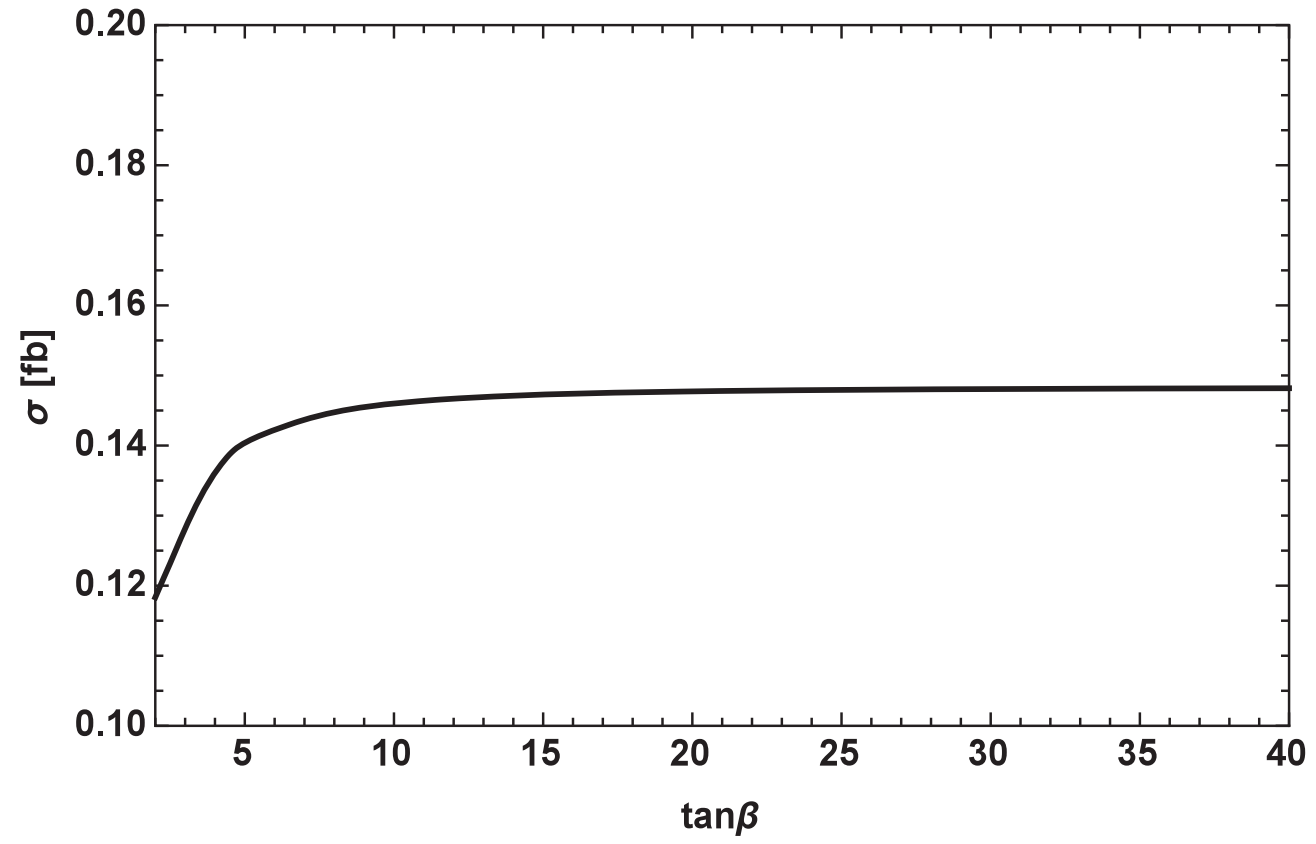

FIG. 8. The diagrams of cross section $\sigma$ versus $\tan\beta$ in the MSSM, where $\mu = -1$ TeV, $E_{\rm cm} = 500$ GeV, $m_{\tilde{L},33} = 100$ GeV, $g_{YB} = 0$, $g_B = 0$, $\tan\beta' = 0$, $Y_{x,33} = 0$, $Y_{\nu,33} = 0$.

In order to further analyze how the new parameters $g_{YB}$ and $g_B$ in the B-LSSM affect the cross section $\sigma$, we plot the Fig. 6, and the gray band also represents the SM prediction with 3 standard deviations [58]. Taking $E_{\rm cm} = 500$ GeV, $\tan\beta' = 1.1$, $m_{\tilde{\nu},33} = 600$ GeV, $m_{\tilde{L},33} = 600$ GeV, $Y_{\nu,33} = 5 \times 10^{-4}$, and $Y_{x,33} = 0.8$, we plot the total cross section of $e^+e^- \to hhZ$ versus the new parameters $g_{YB}$ and $g_B$ in the Fig. 6. In the Fig. 6(a), we take $\tan\beta = 38$, and $g_B = 0.7$ (solid line), $g_B = 0.5$ (dashed line), $g_B = 0.3$ (dotted line), respectively. In the Fig. 6(b), we take $g_{YB} = -0.7$ and $\tan\beta = 35$ (solid line), $\tan\beta = 25$ (dashed line), $\tan\beta = 15$ (dotted line), respectively. The total cross section $\sigma$ in the B-LSSM can exceed that in the SM easily when $g_B$ and $|g_{YB}|$ is small. In addition, the theoretical prediction on the total cross section $\sigma$ depends on the new parameters $g_B$ and $g_{YB}$ strongly. In the Fig. 6(a), with the decreasing of $|g_{YB}|$, the total cross section increases sharply. In the Fig. 6(b), the total cross section $\sigma$ decreases steeply when $g_B$ increases.

Taking $E_{\rm cm} = 500$ GeV, $\tan\beta' = 1.1$, $m_{\tilde{\nu},33} = 600$ GeV, $m_{\tilde{L},33} = 600$ GeV, $g_B = 0.4$, $\tan\beta = 25$, $Y_{\nu,33} = 5 \times 10^{-4}$, we plot the Fig. 7, where the gray band represents the SM prediction with 3 standard deviations [58]. In the Fig. 7, we take $g_{YB} = -0.9$ (solid line), $g_{YB} = -0.7$ (dashed line), $g_{YB} = -0.5$ (dotted line), respectively. Obviously, the dependence of total cross section on Yukawa coupling $Y_{x,33}$ is mild, and, with the decreasing of $|g_{YB}|$, the total cross section increases. The mass of sneutrino in the one-loop effective potential is much smaller than the mass of the stop quark and can be almost ignored, so $Y_{x,33}$ has a small effect on the total cross section $\sigma$. In addition, $Y_{\nu,33}$ is in the order of $10^{-4}$, so it also has little effect on the result.

We also plot the figure with the total cross section $\sigma$ of $e^+e^- \to hhZ$ versus the parameter $\tan\beta$ in the MSSM. In order to compare with the MSSM in Ref. [59], we adopt $\mu = -1$ TeV, $E_{\rm cm} = 500$ GeV, $m_{\tilde{L},33} = 100$ GeV, $g_{YB} = 0$, $g_B = 0$, $\tan\beta' = 0$, $Y_{x,33} = 0$, $Y_{\nu,33} = 0$ to draw

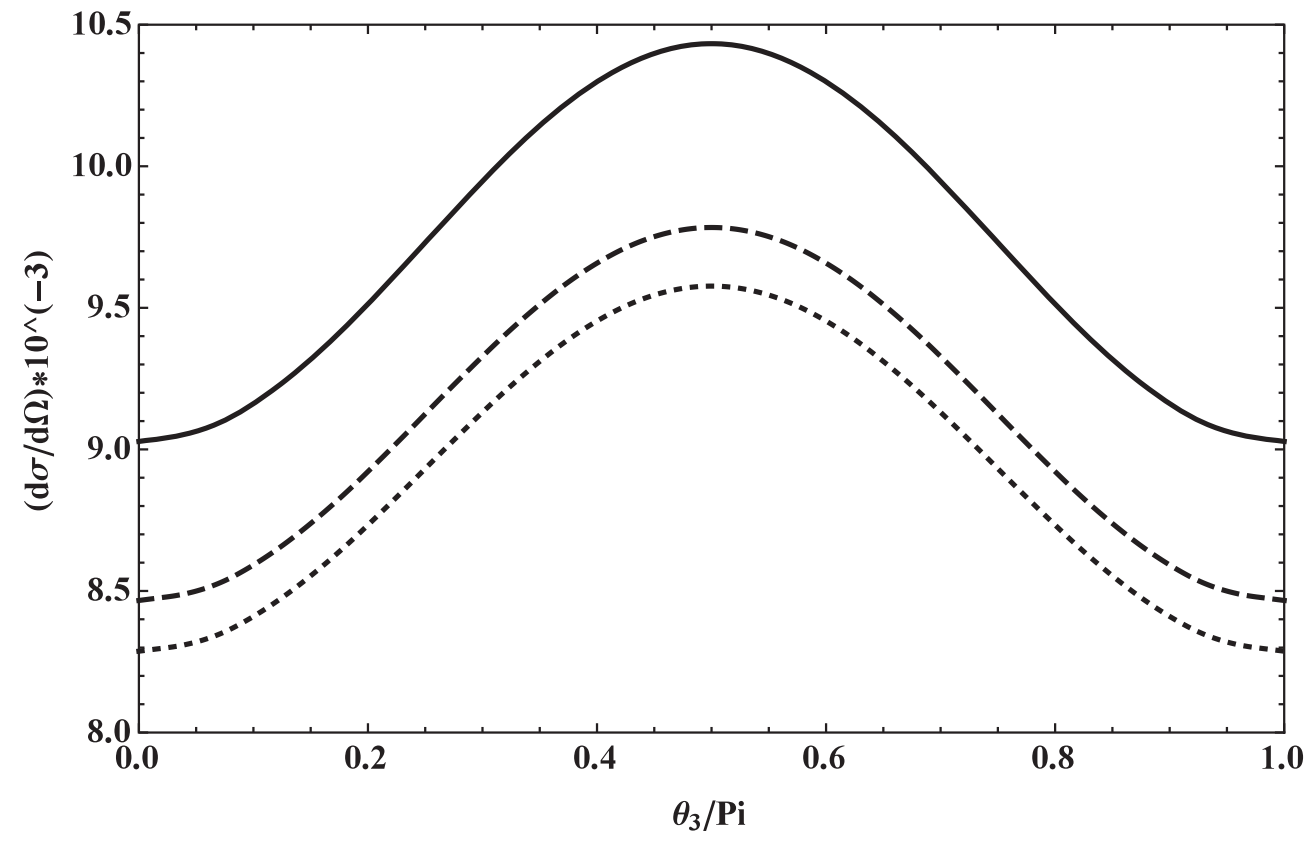

FIG. 9. The picture of differential cross section $\frac{d\sigma}{d\Omega}$ versus angle distribution $\theta_3$, where the solid line denotes $\tan\beta = 18$, the dashed line denotes $\tan\beta = 28$, and the dotted line denotes $\tan\beta = 38$.

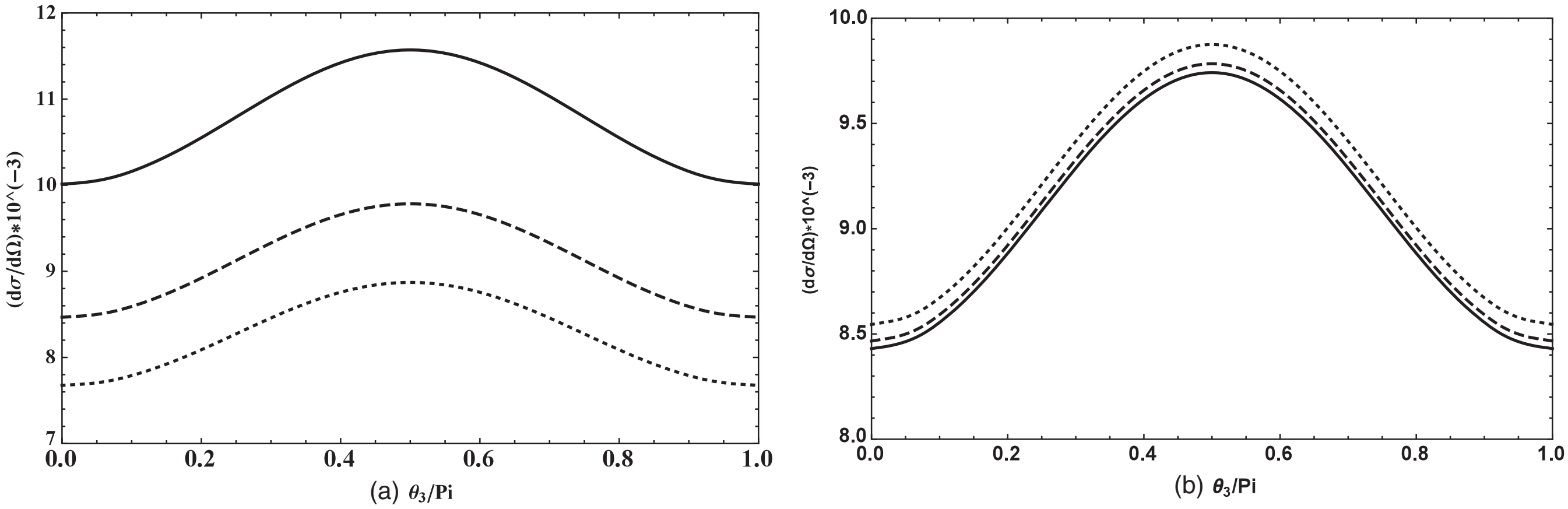

FIG. 10. The picture of differential cross section $\frac{d\sigma}{d\Omega}$ versus angle distribution $\theta_3$, where (a) $g_{YB} = -0.2$ (solid line), $g_{YB} = -0.3$ (dashed line), $g_{YB} = -0.4$ (dotted line), and (b) $g_B = 0.7$ (solid line), $g_B = 0.5$ (dashed line), $g_B = 0.3$ (dotted line).

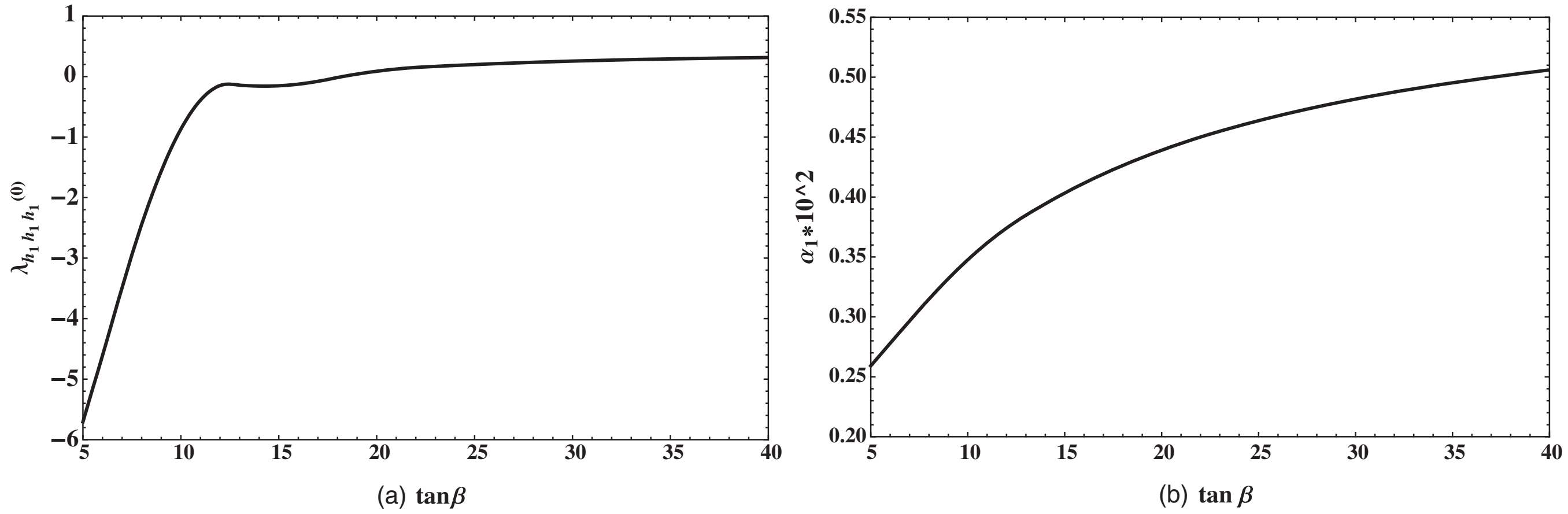

FIG. 11. The picture of $\lambda^{(0)}_{h_1h_1h_1}$ and $\alpha_1$ versus $\tan\beta$, where (a) $\tan\beta' = 1.06$, $g_B = 0.35$, and $g_{YB} = -0.4$, and (b) $\tan\beta' = 1.09$, $g_B = 0.4$, $g_{YB} = -0.4$, $Y_{\nu,33} = 5\times10^{-4}$, and $Y_{x,33} = 0.8$.

Fig. 8. Along with the increasing of $\tan\beta$, the total cross section increases as $\tan\beta \le 5$. When $\tan\beta > 10$, the dependence of total cross section on $\tan\beta$ is mild. In addition, the total cross section is always within the prediction range of SM.

Furthermore, we analyze the influence of the parameters on the angular distribution of differential cross section. Taking $E_{\rm cm} = 500$ GeV, $\tan\beta' = 1.1$, $m_{\tilde{\nu},33} = 600$ GeV, $m_{\tilde{L},33} = 600$ GeV, $Y_{\nu,33} = 5\times10^{-4}$, and $Y_{x,33} = 0.8$, we plot the Figs. 9 and 10. In the Fig. 9, where $g_B = 0.5$, $g_{YB} = -0.3$, and $\tan\beta = 18$ (solid line), $\tan\beta = 28$ (dashed line), $\tan\beta = 38$ (dotted line), respectively. In the Fig. 10(a), we take $\tan\beta = 28$, $g_B = 0.5$ and $g_{YB} = -0.2$ (solid line), $g_{YB} = -0.3$ (dashed line), $g_{YB} = -0.4$ (dotted line), respectively. In the Fig. 10(b), we take $\tan\beta = 28$, $g_{YB} = -0.3$ and $g_B = 0.7$ (solid line), $g_B = 0.5$ (dashed line), $g_B = 0.3$ (dotted line), respectively.

Obviously, when $\theta_3 = \frac{\pi}{2}$, the differential cross section reaches maximum, and when $\theta_3 = 0$ and $\pi$, the differential cross section reaches minimum. The theoretical prediction on the differential cross section depends on the parameters $\tan\beta$, $g_{YB}$, and $g_B$ strongly, along with when the parameters $\tan\beta$, $|g_{YB}|$, and $g_B$ decrease, the differential cross section increases.

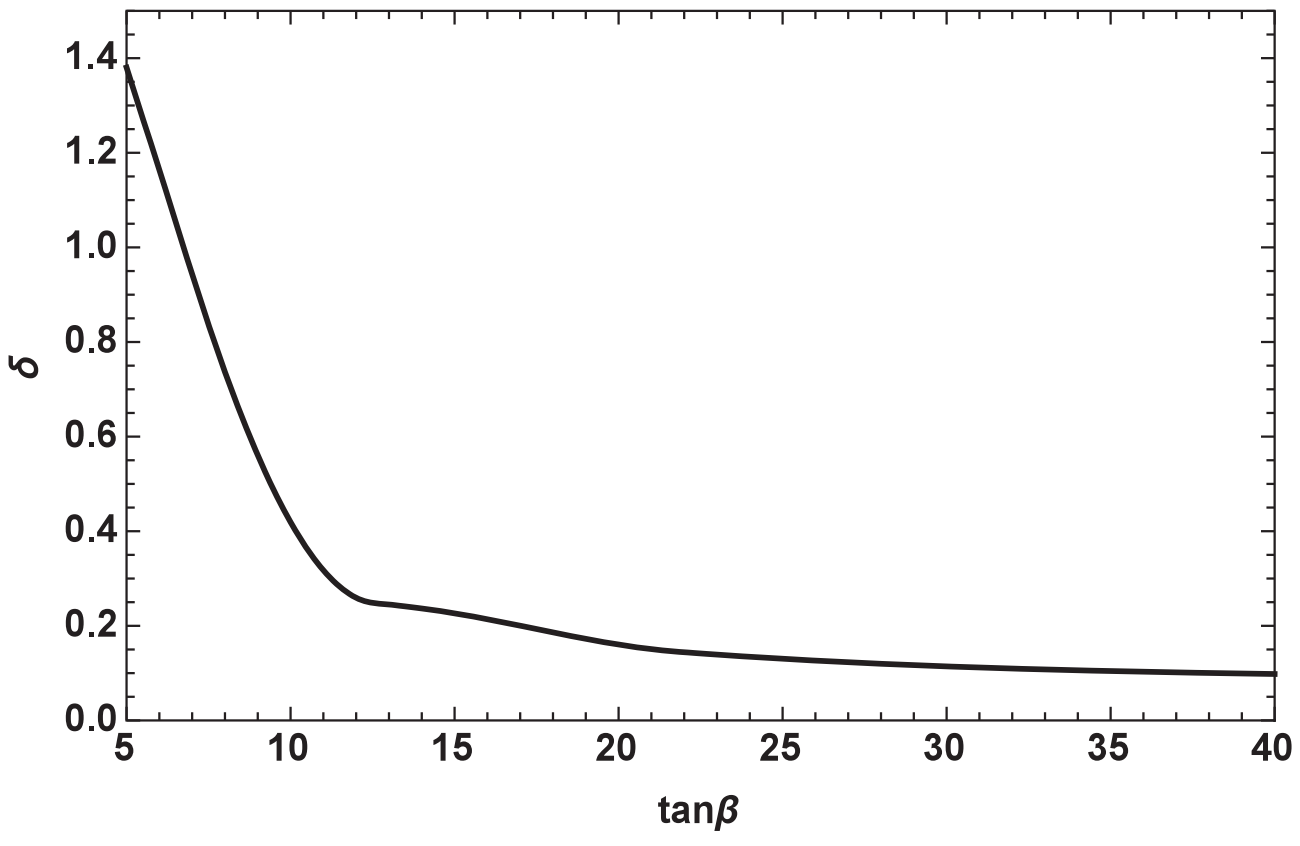

FIG. 12. The picture of $\delta$ versus $\tan\beta$, where $E_{\rm cm} = 500$ GeV, $\tan\beta' = 1.1$, $g_B = 0.4$, $g_{YB} = -0.4$, $Y_{\nu,33} = 5\times10^{-4}$, and $Y_{x,33} = 0.8$.

Now we analyze the self-coupling of Higgs, where $\lambda^{(0)}_{h_1h_1h_1}$ denotes the tree level contribution of triple Higgs self-coupling, and $\alpha_1 = \frac{\lambda_{h_1h_1h_1} - \lambda^{(0)}_{h_1h_1h_1}}{\lambda^{(0)}_{h_1h_1h_1}}$. Taking $\tan\beta' = 1.06$, $g_B = 0.35$, and $g_{YB} = -0.4$, we plot the dependence of self-coupling on the parameter $\tan\beta$ in the Fig. 11(a). Obviously, $\lambda^{(0)}_{h_1h_1h_1}$ increases steeply as $5 < \tan\beta < 10$. Meanwhile, the dependence of self-coupling on the parameter $\tan\beta$ is mild when $\tan\beta > 15$. Taking $\tan\beta' = 1.09$, $g_B = 0.4$, $g_{YB} = -0.4$, $m_{\tilde{\nu},33} = 600$ GeV, $m_{\tilde{L},33} = 600$ GeV, $Y_{\nu,33} = 5\times10^{-4}$, and $Y_{x,33} = 0.8$, we plot dependence of the relative corrections of self-coupling from the one-loop effective potential varying with $\tan\beta$ accordingly. Obviously the relative correction increases steeply when $\tan\beta$ increases.

Finally we present the relative correction on the production cross section $\delta = \frac{\sigma - \sigma_0}{\sigma_0}$ versus the parameter $\tan\beta$, where $\sigma_0$ denotes the theoretical prediction of the total cross section in the SM. Taking $E_{\rm cm} = 500$ GeV, $\tan\beta' = 1.1$, $g_B = 0.4$, $g_{YB} = -0.4$, $m_{\tilde{\nu},33} = 600$ GeV, $m_{\tilde{L},33} = 600$ GeV, $Y_{\nu,33} = 5\times10^{-4}$, and $Y_{x,33} = 0.8$, we plot the relative correction versus $\tan\beta$ in the Fig. 12. The theoretical prediction on the production cross section of the B-LSSM deviates that of the SM obviously as $\tan\beta < 10$. The dependence of the relative correction on $\tan\beta$ changes mildly as $\tan\beta > 25$.

## V. CONCLUSION

In this work we analyze the production cross section of $e^+e^- \to hhZ$ and the self-coupling of Higgs in the B-LSSM. Some parameters affect the theoretical prediction on the production cross section of $e^+e^- \to hhZ$ strongly, for example the new gauge coupling $g_{YB}$. Actually the theoretical prediction on the cross section deviates from that of the SM obviously under some assumptions on the parameters of the model. Nevertheless, the correction from one-loop effective potential to the self-coupling of the lightest Higgs can be neglected safely.

Although the cross section has not been measured when the center-of-mass energy is 500 GeV, the trilinear Higgs boson coupling can be measured at international linear collider through this process. In the future, if the experimental value greatly exceeds the SM, the model can be used to explain the deviation. If the experimental value is consistent with the SM, it will constrain our parameter space.

## ACKNOWLEDGMENTS

We are very grateful to Shu-min Zhao the professor of Hebei University, for giving us some useful discussions. This work is supported by National Natural Science Foundation of China (NNSFC) (Grants No. 11535002, No. 11605037, and No. 11705045), Hebei Key Lab of Optic-Electronic Information and Materials, the midwest universities comprehensive strength promotion project and the youth top-notch talent support program of the Hebei Province, Advanced Talents Incubation Program of the Hebei University, 521000981396.

## APPENDIX A: THE ONE LOOP CORRECTION TO MASS SQUARED MATRIX FOR *CP*-EVEN HIGGS BOSONS

The detailed expression about stability condition:

$$\begin{aligned}
\frac{\partial V}{\partial \mathrm{ReH}_1^1} = {} & -v_2 B\mu + \frac{1}{8}(2g_{YB}g_B u^2 \cos 2\beta' + g^2 v^2 \cos 2\beta) v_1 + (m_{H_1}^2 + \mu^2) v_1 \\
& + \frac{3}{16\pi^2} f(Q^2, m_{\tilde{t}_1}^2)\left[\frac{1}{8} g^2 v_1 + \frac{1}{m_{\tilde{t}_1}^2 - m_{\tilde{t}_2}^2}\left(-\frac{1}{48} M_{tm}^2 C_1^2 v_1 + \sqrt{2} F_4\right)\right] \\
& + \frac{3}{16\pi^2} f(Q^2, m_{\tilde{t}_2}^2)\left[\frac{1}{8} g^2 v_1 - \frac{1}{m_{\tilde{t}_1}^2 - m_{\tilde{t}_2}^2}\left(-\frac{1}{48} M_{tm}^2 C_1^2 v_1 + \sqrt{2} F_4\right)\right] \\
& + \frac{3}{16\pi^2} f(Q^2, m_{\tilde{b}_1}^2)\left[\frac{1}{8} g^2 v_2 + \frac{1}{m_{\tilde{b}_1}^2 - m_{\tilde{b}_2}^2}\left(-\frac{1}{48} M_{bm}^2 C_1^2 v_2 + \sqrt{2} F_4'\right)\right] \\
& + \frac{3}{16\pi^2} f(Q^2, m_{\tilde{b}_2}^2)\left[\frac{1}{8} g^2 v_2 - \frac{1}{m_{\tilde{b}_1}^2 - m_{\tilde{b}_2}^2}\left(-\frac{1}{48} M_{bm}^2 C_1^2 v_2 + \sqrt{2} F_4'\right)\right] \\
& - \frac{3}{16\pi^2} f(Q^2, m_b^2) Y_b m_b + \frac{1}{16\pi^2} f(Q^2, m_{\tilde{\nu}_R}^2)\left\{\frac{1}{8} g^2 v_1 - \frac{1}{4} Y_1^{-\frac{1}{2}}\left[\frac{1}{2}(m_{\tilde{\nu}_{LL}^I}^2 + m_{\tilde{\nu}_{RR}^I}^2) g^2 v_1\right.\right. \\
& \left.\left. - (g^2 + g_{YB}g_B) v_1 m_{\tilde{\nu}_{RR}^I}^2 - g_{YB}g_B v_1 m_{\tilde{\nu}_{LL}^I}^2 - \sqrt{2}\mu Y_{\nu,ij} m_{\tilde{\nu}_{LR}^I}^2\right]\right\}, \qquad \text{(A1)}
\end{aligned}$$

$$\frac{\partial V}{\partial \mathrm{ReH}_2^2} = \frac{1}{8}(2g_{YB}g_B u^2\cos 2\beta' + g^2v^2\cos 2\beta)v_2 - v_1 B\mu + v_2(m_{H_2}^2+\mu^2)$$
$$+\frac{3}{16\pi^2}f(Q^2,m_{\tilde{t}_1}^2)\left[-\frac{1}{8}g^2v_2+\frac{1}{2}Y_t^2v_2+\frac{1}{m_{\tilde{t}_1}^2-m_{\tilde{t}_2}^2}\left(\frac{1}{48}M_{tm}^2C_1^2v_2+\sqrt{2}F_2\right)\right]$$
$$+\frac{3}{16\pi^2}\left\{f(Q^2,m_{\tilde{t}_2}^2)\left[-\frac{1}{8}g^2v_2+\frac{1}{2}Y_t^2v_2-\frac{1}{m_{\tilde{t}_1}^2-m_{\tilde{t}_2}^2}\left(\frac{1}{48}M_{tm}^2C_1^2v_2+\sqrt{2}F_2\right)\right]-f(Q^2,m_t^2)Y_tm_t\right\}$$
$$+\frac{3}{16\pi^2}f(Q^2,m_{\tilde{b}_1}^2)\left[-\frac{1}{8}g^2v_1+\frac{1}{2}Y_b^2v_1+\frac{1}{m_{\tilde{b}_1}^2-m_{\tilde{b}_2}^2}\left(\frac{1}{48}M_{bm}^2C_1^2v_1+\sqrt{2}F_2'\right)\right]$$
$$+\frac{3}{16\pi^2}f(Q^2,m_{\tilde{b}_2}^2)\left[-\frac{1}{8}g^2v_1+\frac{1}{2}Y_b^2v_1-\frac{1}{m_{\tilde{b}_1}^2-m_{\tilde{b}_2}^2}\left(\frac{1}{48}M_{bm}^2C_1^2v_1+\sqrt{2}F_2'\right)\right]$$
$$+\frac{1}{16\pi^2}f(Q^2,m_{\tilde{\nu}_R}^2)\left\{\left(-\frac{1}{8}g^2+Y_{\nu,ij}^2\right)v_2-\frac{1}{4}Y_1^{-\frac{1}{2}}\left[\left(-\frac{1}{8}g^2+Y_{\nu,ij}^2\right)(m_{\tilde{\nu}_{LL}^I}^2+m_{\tilde{\nu}_{RR}^I}^2)v_2\right.\right.$$
$$\left.\left.+(g^2+g_{YB}g_B-4Y_{\nu,ij}^2)v_2m_{\tilde{\nu}_{RR}^I}^2-(g_{YB}g_B+4Y_{\nu,ij}^2)v_2m_{\tilde{\nu}_{LL}^I}^2+(\sqrt{2}T_\nu^{ij}+2u_1Y_{\nu,ij}Y_{x,ij})m_{\tilde{\nu}_{LR}^I}^2\right]\right\}, \tag{A2}$$

$$\frac{\partial V}{\partial \mathrm{Re}\tilde{\eta}_1} = \frac{1}{4}(2g_B^2u^2\cos 2\beta' + g_{YB}g_Bv^2\cos 2\beta)u_1 + (m_{\tilde{\eta}_1}^2+\mu^2)u_1 - u_2B\mu + \frac{3}{64\pi^2}f(Q^2,m_{\tilde{t}_1}^2)\left[g_{YB}g_Bu_1+\frac{M_{tm}^2C_3^2u_1}{12(m_{\tilde{t}_1}^2-m_{\tilde{t}_2}^2)}\right]$$
$$+\frac{3}{64\pi^2}f(Q^2,m_{\tilde{t}_2}^2)\left[g_{YB}g_Bu_1-\frac{M_{tm}^2C_3^2u_1}{12(m_{\tilde{t}_1}^2-m_{\tilde{t}_2}^2)}\right]+\frac{3}{64\pi^2}f(Q^2,m_{\tilde{b}_1}^2)\left[g_{YB}g_Bu_2+\frac{M_{bm}^2C_3^2u_2}{12(m_{\tilde{b}_1}^2-m_{\tilde{b}_2}^2)}\right]$$
$$+\frac{3}{64\pi^2}f(Q^2,m_{\tilde{b}_2}^2)\left[g_{YB}g_Bu_2-\frac{M_{bm}^2C_3^2u_2}{12(m_{\tilde{b}_1}^2-m_{\tilde{b}_2}^2)}\right]$$
$$+\frac{1}{16\pi^2}\left\{f(Q^2,m_{\tilde{\nu}_R}^2)\left\{\frac{1}{4}g_{YB}g_Bu_1+\frac{\sqrt{2}}{2}T_x^{ij}+2u_1Y_{x,ij}^2-\frac{1}{4}Y_1^{-\frac{1}{2}}[(g_{YB}g_Bu_1+2\sqrt{2}T_x^{ij}+8u_1Y_{x,ij}^2)(m_{\tilde{\nu}_{LL}^I}^2+m_{\tilde{\nu}_{RR}^I}^2)\right.\right.$$
$$\left.\left.-2(g_{YB}g_B+g_B^2)u_1m_{\tilde{\nu}_{RR}^I}^2-m_{\tilde{\nu}_{LL}^I}^2(-2g_B^2u_1+4\sqrt{2}T_x^{ij}+16u_1Y_{x,ij}^2)]\right\}-8f(Q,m_{\nu_R}^2)u_1Y_{x,ij}^2\right\}, \tag{A3}$$

$$\frac{\partial V}{\partial \mathrm{Re}\tilde{\eta}_2} = \frac{1}{4}(g_{YB}g_Bv^2\cos 2\beta - 2g_B^2u^2\cos 2\beta')u_2 + u_2(m_{\tilde{\eta}_2}^2+\mu^2) - u_1B\mu + \frac{3}{64\pi^2}f(Q^2,m_{\tilde{t}_1}^2)\left[-g_{YB}g_Bu_2-\frac{M_{tm}^2C_3^2u_2}{12(m_{\tilde{t}_1}^2-m_{\tilde{t}_2}^2)}\right]$$
$$+\frac{3}{64\pi^2}f(Q^2,m_{\tilde{t}_2}^2)\left[-g_{YB}g_Bu_2+\frac{M_{tm}^2C_3^2u_2}{12(m_{\tilde{t}_1}^2-m_{\tilde{t}_2}^2)}\right]$$
$$+\frac{3}{64\pi^2}f(Q^2,m_{\tilde{b}_1}^2)\left[-g_{YB}g_Bu_1-\frac{M_{bm}^2C_3^2u_1}{12(m_{\tilde{b}_1}^2-m_{\tilde{b}_2}^2)}\right]$$
$$+\frac{3}{64\pi^2}f(Q^2,m_{\tilde{b}_2}^2)\left[-g_{YB}g_Bu_1+\frac{M_{bm}^2C_3^2u_1}{12(m_{\tilde{b}_1}^2-m_{\tilde{b}_2}^2)}\right]$$
$$+\frac{1}{64\pi^2}f(Q^2,m_{\tilde{\nu}_R}^2)\{-g_{YB}g_Bu_2-2\sqrt{2}\mu Y_{x,ij}-Y_1^{-\frac{1}{2}}[(-g_B^2u_2-g_{YB}g_Bu_2+g_B^2u_1-\sqrt{2}\mu Y_{x,ij})(m_{\tilde{\nu}_{LL}^I}^2+m_{\tilde{\nu}_{RR}^I}^2)$$
$$-(2g_B^2+2g_{YB}g_B)u_2m_{\tilde{\nu}_{RR}^I}^2-m_{\tilde{\nu}_{LL}^I}^2(2g_B^2u_2-4\sqrt{2}\mu Y_{x,ij})]\}. \tag{A4}$$

The detailed expression about $\Delta m_h^2$:

$$\Delta m^2_{\Phi_d\Phi_d} = \frac{3}{16\pi^2}\Bigg\{[f(Q^2,m^2_{\tilde{t}_1}) - f(Q^2,m^2_{\tilde{t}_2})]\Bigg[-\frac{C_1^4 v_1^2}{576(m^2_{\tilde{t}_1}-m^2_{\tilde{t}_2})} - \frac{v_1^2(48F_5 - C_1^2 M^2_{tm})^2}{1152(m^2_{\tilde{t}_1}-m^2_{\tilde{t}_2})^3}\Bigg]$$
$$+\frac{1}{4608}\ln\frac{m^2_{\tilde{t}_1}}{Q^2}\Bigg[(24Y_t^2 - C_2^2)v_1 + \frac{48F_4 - C_1^2 M^2_{tm} v_1}{m^2_{\tilde{t}_1}-m^2_{\tilde{t}_2}}\Bigg]^2$$
$$+\frac{1}{4608}\ln\frac{m^2_{\tilde{t}_2}}{Q^2}\Bigg[(24Y_t^2 - C_2^2)v_1 - \frac{48F_4 - C_1^2 M^2_{tm} v_1}{m^2_{\tilde{t}_1}-m^2_{\tilde{t}_2}}\Bigg]^2\Bigg\}$$
$$+\frac{3}{16\pi^2}\Bigg\{[f(Q^2,m^2_{\tilde{b}_1}) - f(Q^2,m^2_{\tilde{b}_2})]\Bigg[-\frac{C_1^4 v_2^2}{576(m^2_{\tilde{b}_1}-m^2_{\tilde{b}_2})} - \frac{v_2^2(48F'_5 - C_1^2 M^2_{bm})^2}{1152(m^2_{\tilde{b}_1}-m^2_{\tilde{b}_2})^3}\Bigg]$$
$$+\frac{1}{4608}\ln\frac{m^2_{\tilde{b}_1}}{Q^2}\Bigg[(24Y_b^2 - C_2'^2)v_2 + \frac{48F'_4 - C_1^2 M^2_{bm} v_2}{m^2_{\tilde{b}_1}-m^2_{\tilde{b}_2}}\Bigg]^2$$
$$+\frac{1}{4608}\ln\frac{m^2_{\tilde{b}_2}}{Q^2}\Bigg[(24Y_b^2 - C_2'^2)v_2 - \frac{48F'_4 - C_1^2 M^2_{bm} v_2}{m^2_{\tilde{b}_1}-m^2_{\tilde{b}_2}}\Bigg]^2 - \frac{1}{4}f(Q^2,m_b^2)Y_b^2\Bigg\}$$
$$+\frac{1}{32\pi^2}\Bigg\{\frac{1}{8}g^2 f(Q^2,m^2_{\tilde{\nu}_R}) + f(Q^2,m^2_{\tilde{\nu}_R})\Bigg\{\frac{1}{8}Y_1^{-\frac{3}{2}}y_d\Bigg[\left(\frac{1}{2}g^2 - g_{YB}g_B\right)v_1 m^2_{\tilde{\nu}^I_{LL}} - \left(\frac{1}{2}g^2 + g_{YB}g_B\right)v_1 m^2_{\tilde{\nu}^I_{RR}} - \sqrt{2}\mu Y_{\nu,ij} m^2_{\tilde{\nu}^I_{LR}}\Bigg]$$
$$+\frac{1}{4}Y_1^{-\frac{1}{2}}\Bigg[\left(\frac{1}{2}g^2 - g_{YB}g_B\right)(m^2_{\tilde{\nu}^I_{LL}} + m^2_{\tilde{\nu}^I_{RR}}) + \frac{1}{8}g^4 v_1^2 - g^2 m^2_{\tilde{\nu}^I_{RR}} - \frac{1}{2}(g^2 + g_{YB}g_B)g_{YB}g_B v_1^2 + \mu^2 Y^2_{\nu,ij}\Bigg]\Bigg\}$$
$$+\frac{1}{4}\ln\frac{m^2_{\tilde{\nu}_R}}{Q^2}y^2_{d_R}\Bigg\}, \tag{A5}$$

$$\Delta m^2_{\Phi_u\Phi_d} = \frac{3}{16\pi^2}\Bigg\{[f(Q^2,m^2_{\tilde{t}_1}) - f(Q^2,m^2_{\tilde{t}_2})]\Bigg[-\frac{C_1^4 v_2 v_1}{1152(m^2_{\tilde{t}_1}-m^2_{\tilde{t}_2})} + \frac{M^2_{tm}C_1^2 v_1 - F_4}{1152(m^2_{\tilde{t}_1}-m^2_{\tilde{t}_2})^3}(M^2_{tm}C_1^2 v_2 + F_2)\Bigg]$$
$$+\frac{1}{96}\ln\frac{m^2_{\tilde{t}_1}}{Q^2}\Bigg[(24Y_t^2 - C_2^2)v_1 + \frac{48F_4 - C_1^2 M^2_{tm} v_1}{48(m^2_{\tilde{t}_1}-m^2_{\tilde{t}_2})}\left(C_2^2 v_2 + \frac{48F_2 + C_1^2 M^2_{tm} v_2}{m^2_{\tilde{t}_1}-m^2_{\tilde{t}_2}}\right)\Bigg]$$
$$+\frac{1}{96}\ln\frac{m^2_{\tilde{t}_2}}{Q^2}\Bigg[(24Y_t^2 - C_2^2)v_1 - \frac{48F_4 - C_1^2 M^2_{tm} v_1}{48(m^2_{\tilde{t}_1}-m^2_{\tilde{t}_2})}\left(C_2^2 v_2 - \frac{48F_2 + C_1^2 M^2_{tm} v_2}{m^2_{\tilde{t}_1}-m^2_{\tilde{t}_2}}\right)\Bigg]\Bigg\}$$
$$+\frac{3}{16\pi^2}\Bigg\{[f(Q^2,m^2_{\tilde{b}_1}) - f(Q^2,m^2_{\tilde{b}_2})]\Bigg[-\frac{C_1^4 v_2 v_1}{1152(m^2_{\tilde{b}_1}-m^2_{\tilde{b}_2})} + \frac{M^2_{bm}C_1^2 v_2 - F'_4}{1152(m^2_{\tilde{b}_1}-m^2_{\tilde{b}_2})^3}(M^2_{bm}C_1^2 v_1 + F'_2)\Bigg]$$
$$+\frac{1}{96}\ln\frac{m^2_{\tilde{b}_1}}{Q^2}\Bigg[(24Y_b^2 - C_2'^2)v_2 + \frac{48F'_4 - C_1^2 M^2_{bm} v_2}{48(m^2_{\tilde{b}_1}-m^2_{\tilde{b}_2})}\left(C_2'^2 v_1 + \frac{48F'_2 + C_1^2 M^2_{bm} v_1}{m^2_{\tilde{b}_1}-m^2_{\tilde{b}_2}}\right)\Bigg]$$
$$+\frac{1}{96}\ln\frac{m^2_{\tilde{b}_2}}{Q^2}\Bigg[(24Y_b^2 - C_2'^2)v_2 - \frac{48F'_4 - C_1^2 M^2_{bm} v_2}{48(m^2_{\tilde{b}_1}-m^2_{\tilde{b}_2})}\left(C_2'^2 v_1 - \frac{48F'_2 + C_1^2 M^2_{bm} v_1}{m^2_{\tilde{b}_1}-m^2_{\tilde{b}_2}}\right)\Bigg]\Bigg\}$$
$$+\frac{1}{32\pi^2}\Bigg\{f(Q^2,m^2_{\tilde{\nu}_R})\Bigg\{\frac{1}{8}Y_1^{-\frac{3}{2}}y_d y_u - \frac{1}{4}Y_1^{-\frac{1}{2}}\Bigg[-(T^{ij}_\nu + \sqrt{2}u_1 Y_{x,ij}Y_{\nu,ij})\mu Y_{\nu,ij}$$
$$-\left(\frac{1}{4}g^2 + \frac{1}{4}g_{YB}g_B - Y^2_{\nu,ij}\right)g_{YB}g_B v_1 v_2 - (g^2 + g_{YB}g_B)\left(\frac{1}{4}g_{YB}g_B + Y^2_{\nu,ij}\right)v_1 v_2$$
$$+\left(Y^2_{\nu,ij} - \frac{1}{8}g^2 v_1\right)g^2 v_1 v_2\Bigg]\Bigg\} + \frac{1}{4}\ln\frac{m^2_{\tilde{\nu}_R}}{Q^2}y_{d_R}\left(-\frac{1}{8}g^2 v_2 + v_2 Y^2_{\nu,ij} - \frac{1}{4}Y_1^{-\frac{1}{2}}y_u\right)\Bigg\}, \tag{A6}$$

$$
\begin{aligned}
\Delta m^2_{\Phi_\eta\Phi_d} = {} & \frac{3}{16\pi^2}\left\{\frac{1}{1152}[f(Q^2,m^2_{\tilde t_1})-f(Q^2,m^2_{\tilde t_2})]\left[-\frac{C_1^2C_3^2v_1u_1}{m^2_{\tilde t_1}-m^2_{\tilde t_2}}-\frac{M^2_{tm}C_3^2u_1(48F_4-C_1^2M^2_{tm}v_1)}{(m^2_{\tilde t_1}-m^2_{\tilde t_2})^3}\right]\right.\\
& +\frac{1}{9216}\ln\frac{m^2_{\tilde t_1}}{Q^2}\left[(24Y_t^2-C_2^2)v_1+\frac{48F_4-C_1^2M^2_{tm}v_1}{m^2_{\tilde t_1}-m^2_{\tilde t_2}}\right]\times\left(C_4^2+\frac{M^2_{tm}C_3^2}{m^2_{\tilde t_1}-m^2_{\tilde t_2}}\right)u_1\\
& \left.+\frac{1}{9216}\ln\frac{m^2_{\tilde t_2}}{Q^2}\times\left[(24Y_t^2-C_2^2)v_1-\frac{48F_4-C_1^2M^2_{tm}v_1}{m^2_{\tilde t_1}-m^2_{\tilde t_2}}\right]\times\left(C_4^2-\frac{M^2_{tm}C_3^2}{m^2_{\tilde t_1}-m^2_{\tilde t_2}}\right)u_1\right\}\\
& +\frac{3}{16\pi^2}\left\{\frac{1}{1152}[f(Q^2,m^2_{\tilde b_1})-f(Q^2,m^2_{\tilde b_2})]\left[-\frac{C_1^2C_3^2v_2u_2}{m^2_{\tilde b_1}-m^2_{\tilde b_2}}-\frac{M^2_{bm}C_3^2u_2(48F'_4-C_1^2M^2_{bm}v_2)}{(m^2_{\tilde b_1}-m^2_{\tilde b_2})^3}\right]\right.\\
& +\frac{1}{9216}\ln\frac{m^2_{\tilde b_1}}{Q^2}\left[(24Y_b^2-C_2'^2)v_2+\frac{48F'_4-C_1^2M^2_{bm}v_2}{m^2_{\tilde b_1}-m^2_{\tilde b_2}}\right]\times\left(C_4^2+\frac{M^2_{bm}C_3^2}{m^2_{\tilde b_1}-m^2_{\tilde b_2}}\right)u_2\\
& \left.+\frac{1}{9216}\ln\frac{m^2_{\tilde b_2}}{Q^2}\times\left[(24Y_b^2-C_2'^2)v_2-\frac{48F'_4-C_1^2M^2_{bm}v_2}{m^2_{\tilde b_1}-m^2_{\tilde b_2}}\right]\times\left(C_4^2-\frac{M^2_{bm}C_3^2}{m^2_{\tilde b_1}-m^2_{\tilde b_2}}\right)u_2\right\}\\
& +\frac{1}{32\pi^2}\left\{f(Q^2,m^2_{\tilde\nu_R})\left\{\frac{1}{8}Y_1^{-\frac{3}{2}}y_dy_\eta-\frac{1}{4}Y_1^{-\frac{1}{2}}\left[\left(\frac{1}{4}g_{YB}g_Bu_1+\frac{\sqrt{2}}{2}T_x^{ij}+2u_1Y^2_{x,ij}\right)g^2v_1\right.\right.\right.\\
& -(g^2+g_{YB}g_B)\left(-\frac{1}{2}g_B^2u_1+\sqrt{2}T_x^{ij}+4u_1Y^2_{x,ij}\right)v_1-\frac{1}{2}(g_B^2+g_{YB}g_B)g_{YB}g_Bv_1u_1\\
& \left.\left.\left.-\sqrt{2}\mu Y^2_{\nu,ii}Y_{x,ij}v_2\right]\right\}+\ln\frac{m^2_{\tilde\nu_R}}{Q^2}y_{d_R}\times\left(\frac{1}{16}g_{YB}g_Bu_1+\frac{\sqrt{2}}{4}T_x^{ij}+u_1Y^2_{x,ij}-\frac{1}{16}Y_1^{-\frac{1}{2}}y_\eta\right)\right\},
\end{aligned}
\tag{A7}
$$

$$
\begin{aligned}
\Delta m^2_{\Phi_{\bar\eta}\Phi_d} = {} & \frac{3}{16\pi^2}\left\{[f(Q^2,m^2_{\tilde t_1})-f(Q^2,m^2_{\tilde t_2})]\left[\frac{C_1^2C_3^2v_1u_2}{1152(m^2_{\tilde t_1}-m^2_{\tilde t_2})}+\frac{M^2_{tm}v_1C_3^2}{24(m^2_{\tilde t_1}-m^2_{\tilde t_2})^3}\left(-\frac{C_1^2}{48}M^2_{tm}u_2+F_4\right)\right]\right.\\
& -\frac{1}{192}\ln\frac{m^2_{\tilde t_1}}{Q^2}\times\left(C_4^2+\frac{M^2_{tm}C_3^2}{m^2_{\tilde t_1}-m^2_{\tilde t_2}}\right)u_2\times\left[\frac{1}{48}(24Y_t^2-C_2^2)v_1+\frac{1}{m^2_{\tilde t_1}-m^2_{\tilde t_2}}\left(-\frac{C_1^2}{48}M^2_{tm}v_1+F_4\right)\right]\\
& \left.+\frac{1}{192}\ln\frac{m^2_{\tilde t_2}}{Q^2}\times\left(-C_4^2+\frac{M^2_{tm}C_3^2}{m^2_{\tilde t_1}-m^2_{\tilde t_2}}\right)u_2\left[\frac{1}{48}(24Y_t^2-C_2^2)v_1-\frac{1}{m^2_{\tilde t_1}-m^2_{\tilde t_2}}\left(-\frac{C_1^2}{48}M^2_{tm}v_1+F_4\right)\right]\right\}\\
& +\frac{3}{16\pi^2}\left\{[f(Q^2,m^2_{\tilde b_1})-f(Q^2,m^2_{\tilde b_2})]\left[\frac{C_1^2C_3^2v_2u_1}{1152(m^2_{\tilde b_1}-m^2_{\tilde b_2})}+\frac{M^2_{bm}v_2C_3^2}{24(m^2_{\tilde b_1}-m^2_{\tilde b_2})^3}\left(-\frac{C_1^2}{48}M^2_{bm}u_1+F'_4\right)\right]\right.\\
& -\frac{1}{192}\ln\frac{m^2_{\tilde b_1}}{Q^2}\times\left(C_4^2+\frac{M^2_{bm}C_3^2}{m^2_{\tilde b_1}-m^2_{\tilde b_2}}\right)u_1\times\left[\frac{1}{48}(24Y_b^2-C_2'^2)v_2+\frac{1}{m^2_{\tilde b_1}-m^2_{\tilde b_2}}\left(-\frac{C_1^2}{48}M^2_{bm}v_2+F'_4\right)\right]\\
& \left.+\frac{1}{192}\ln\frac{m^2_{\tilde b_2}}{Q^2}\times\left(-C_4^2+\frac{M^2_{bm}C_3^2}{m^2_{\tilde b_1}-m^2_{\tilde b_2}}\right)u_1\left[\frac{1}{48}(24Y_b^2-C_2'^2)v_2-\frac{1}{m^2_{\tilde b_1}-m^2_{\tilde b_2}}\left(-\frac{C_1^2}{48}M^2_{bm}v_2+F'_4\right)\right]\right\}\\
& +\frac{1}{32\pi^2}\left\{f(Q^2,m^2_{\tilde\nu_R})\left\{\frac{1}{8}Y_1^{-\frac{3}{2}}y_dy_{\bar\eta}-\frac{1}{4}Y_1^{-\frac{1}{2}}\left[\left(-\frac{1}{4}g_{YB}g_Bu_2-\frac{\sqrt{2}}{2}\mu Y_{x,ij}\right)g^2v_1\right.\right.\right.\\
& \left.\left.-(g^2+g_{YB}g_B)\left(\frac{1}{2}g_B^2u_2-\sqrt{2}\mu Y_{x,ij}\right)v_1-\frac{1}{2}(g_B^2+g_{YB}g_B)g_{YB}g_Bv_1u_2\right]\right\}\\
& \left.+\ln\frac{m^2_{\tilde\nu_R}}{Q^2}y_{d_R}\left(-\frac{1}{16}g_{YB}g_Bu_2-\frac{\sqrt{2}}{8}\mu Y_{x,ij}-\frac{1}{16}Y_1^{-\frac{1}{2}}y_{\bar\eta}\right)\right\},
\end{aligned}
\tag{A8}
$$

$$\Delta m^2_{\Phi_u\Phi_u} = \frac{3}{16\pi^2}\Bigg\{[f(Q^2,m^2_{\tilde t_1}) - f(Q^2,m^2_{\tilde t_2})]\Bigg[\frac{C_1^4 v_2^2}{1152(m^2_{\tilde t_1}-m^2_{\tilde t_2})} - \frac{2v_2^2}{(m^2_{\tilde t_1}-m^2_{\tilde t_2})^3}\times\left(\frac{C_1^2}{48}M^2_{tm}+A_t^2Y_t^2\right)^2\Bigg]$$
$$+\frac{1}{192}\ln\frac{m^2_{\tilde t_1}}{Q^2}\left(C_2^2v_2+\frac{C_1^2M^2_{tm}v_2+48F_2}{m^2_{\tilde t_1}-m^2_{\tilde t_2}}\right)^2+\frac{1}{192}\ln\frac{m^2_{\tilde t_2}}{Q^2}\left(C_2^2v_2-\frac{C_1^2M^2_{tm}v_2+48F_2}{m^2_{\tilde t_1}-m^2_{\tilde t_2}}\right)^2\Bigg\}$$
$$-\frac{1}{4}f(Q^2,m_t^2)Y_t^2+\frac{3}{16\pi^2}\Bigg\{[f(Q^2,m^2_{\tilde b_1})-f(Q^2,m^2_{\tilde b_2})]\Bigg[\frac{C_1^4v_1^2}{1152(m^2_{\tilde b_1}-m^2_{\tilde b_2})}$$
$$-\frac{2v_1^2}{(m^2_{\tilde b_1}-m^2_{\tilde b_2})^3}\times\left(\frac{C_1^2}{48}M^2_{bm}+A_b^2Y_b^2\right)^2\Bigg]+\frac{1}{192}\ln\frac{m^2_{\tilde b_1}}{Q^2}\left(C_2'^2v_1+\frac{C_1^2M^2_{bm}v_1+48F_2'}{m^2_{\tilde b_1}-m^2_{\tilde b_2}}\right)^2$$
$$+\frac{1}{192}\ln\frac{m^2_{\tilde b_2}}{Q^2}\left(C_2'^2v_1-\frac{C_1^2M^2_{bm}v_1+48F_2'}{m^2_{\tilde b_1}-m^2_{\tilde b_2}}\right)^2\Bigg\}+\frac{1}{32\pi^2}\Bigg[f(Q^2,m^2_{\tilde\nu_R})\left(-\frac{1}{8}g^2+Y^2_{\nu,ij}\right)$$
$$+f(Q^2,m^2_{\tilde\nu_R})\left(\frac{1}{8}Y_1^{-\frac{3}{2}}y_u^2-\frac{1}{4}Y_1^{-\frac{1}{2}}y_{u_1}\right)+\frac{1}{4}\ln\frac{m^2_{\tilde\nu_R}}{Q^2}y^2_{u_R}\Bigg], \tag{A9}$$

$$\Delta m^2_{\Phi_\eta\Phi_u} = \frac{3}{16\pi^2}\Bigg\{[f(Q^2,m^2_{\tilde t_1})-f(Q^2,m^2_{\tilde t_2})]\Bigg[\frac{C_3^2C_1^2u_1v_2}{1152(m^2_{\tilde t_1}-m^2_{\tilde t_2})}-\frac{M^2_{tm}C_3^2u_1v_2}{24(m^2_{\tilde t_1}-m^2_{\tilde t_2})^3}\left(\frac{C_1^2}{48}M^2_{tm}+A_t^2Y_t^2\right)\Bigg]$$
$$+\frac{1}{9216}\ln\frac{m^2_{\tilde t_1}}{Q^2}\times\left(C_4^2+\frac{M^2_{tm}C_3^2}{m^2_{\tilde t_1}-m^2_{\tilde t_2}}\right)u_1\times\left(C_2^2v_2+\frac{C_1^2M^2_{tm}v_2+48F_2}{m^2_{\tilde t_1}-m^2_{\tilde t_2}}\right)$$
$$+\frac{1}{9216}\ln\frac{m^2_{\tilde t_2}}{Q^2}\times\left(C_4^2-\frac{M^2_{tm}C_3^2}{m^2_{\tilde t_1}-m^2_{\tilde t_2}}\right)u_1\times\left(C_2^2v_2-\frac{C_1^2M^2_{tm}v_2+48F_2}{m^2_{\tilde t_1}-m^2_{\tilde t_2}}\right)\Bigg\}$$
$$+\frac{3}{16\pi^2}\Bigg\{[f(Q^2,m^2_{\tilde b_1})-f(Q^2,m^2_{\tilde b_2})]\Bigg[\frac{C_3^2C_1^2u_2v_1}{1152(m^2_{\tilde b_1}-m^2_{\tilde b_2})}-\frac{M^2_{bm}C_3^2u_2v_1}{24(m^2_{\tilde b_1}-m^2_{\tilde b_2})^3}\left(\frac{C_1^2}{48}M^2_{bm}+A_b^2Y_b^2\right)\Bigg]$$
$$+\frac{1}{9216}\ln\frac{m^2_{\tilde b_1}}{Q^2}\times\left(C_4^2+\frac{M^2_{bm}C_3^2}{m^2_{\tilde b_1}-m^2_{\tilde b_2}}\right)u_2\times\left(C_2'^2v_1+\frac{C_1^2M^2_{bm}v_1+48F_2'}{m^2_{\tilde b_1}-m^2_{\tilde b_2}}\right)+\frac{1}{9216}\ln\frac{m^2_{\tilde b_2}}{Q^2}\times\left(C_4^2-\frac{M^2_{bm}C_3^2}{m^2_{\tilde b_1}-m^2_{\tilde b_2}}\right)u_2$$
$$\times\left(C_2'^2v_1-\frac{C_1^2M^2_{bm}v_1+48F_2'}{m^2_{\tilde b_1}-m^2_{\tilde b_2}}\right)\Bigg\}+\frac{1}{32\pi^2}\Bigg[f(Q^2,m^2_{\tilde\nu_R})\left(\frac{1}{8}Y_1^{-\frac{3}{2}}y_\eta y_u-\frac{1}{4}Y_1^{-\frac{1}{2}}y_{u_{1\eta}}\right)+\frac{1}{4}\ln\frac{m^2_{\tilde\nu_R}}{Q^2}y_{u_R}y_{\eta_R}\Bigg], \tag{A10}$$

$$\Delta m^2_{\Phi_{\bar\eta}\Phi_u} = \frac{3}{16\pi^2}\Bigg\{[f(Q^2,m^2_{\tilde t_1})-f(Q^2,m^2_{\tilde t_2})]\Bigg[-\frac{C_3^2C_1^2v_2u_2}{1152(m^2_{\tilde t_1}-m^2_{\tilde t_2})}+\frac{M^2_{tm}C_3^2u_2v_2}{24(m^2_{\tilde t_1}-m^2_{\tilde t_2})^3}\left(\frac{C_1^2}{48}M^2_{tm}+F_2\right)\Bigg]$$
$$+\frac{1}{9216}\ln\frac{m^2_{\tilde t_1}}{Q^2}\left(-C_3^2-\frac{M^2_{tm}C_3^2}{m^2_{\tilde t_1}-m^2_{\tilde t_2}}\right)u_2\times\left(C_2^2v_2+\frac{C_1^2M^2_{tm}v_2+48F_2}{m^2_{\tilde t_1}-m^2_{\tilde t_2}}\right)$$
$$+\frac{1}{9216}\ln\frac{m^2_{\tilde t_2}}{Q^2}\left(-C_3^2+\frac{M^2_{tm}C_3^2}{m^2_{\tilde t_1}-m^2_{\tilde t_2}}\right)u_2\times\left(C_2^2v_2-\frac{C_1^2M^2_{tm}v_2+48F_2}{m^2_{\tilde t_1}-m^2_{\tilde t_2}}\right)\Bigg\}$$
$$+\frac{3}{16\pi^2}\Bigg\{[f(Q^2,m^2_{\tilde b_1})-f(Q^2,m^2_{\tilde b_2})]\Bigg[-\frac{C_3^2C_1^2v_1u_1}{1152(m^2_{\tilde b_1}-m^2_{\tilde b_2})}+\frac{M^2_{bm}C_3^2u_1v_1}{24(m^2_{\tilde b_1}-m^2_{\tilde b_2})^3}\left(\frac{C_1^2}{48}M^2_{bm}+F_2'\right)\Bigg]$$
$$+\frac{1}{9216}\ln\frac{m^2_{\tilde b_1}}{Q^2}\left(-C_3^2-\frac{M^2_{bm}C_3^2}{m^2_{\tilde b_1}-m^2_{\tilde b_2}}\right)u_1\times\left(C_2'^2v_1+\frac{C_1^2M^2_{bm}v_1+48F_2'}{m^2_{\tilde b_1}-m^2_{\tilde b_2}}\right)+\frac{1}{9216}\ln\frac{m^2_{\tilde b_2}}{Q^2}\left(-C_3^2+\frac{M^2_{bm}C_3^2}{m^2_{\tilde b_1}-m^2_{\tilde b_2}}\right)u_1$$
$$\times\left(C_2'^2v_1-\frac{C_1^2M^2_{bm}v_1+48F_2'}{m^2_{\tilde b_1}-m^2_{\tilde b_2}}\right)\Bigg\}+\frac{1}{32\pi^2}\Bigg[f(Q^2,m^2_{\tilde\nu_R})\left(\frac{1}{8}Y_1^{-\frac{3}{2}}y_{\bar\eta}y_u-\frac{1}{4}Y_1^{-\frac{1}{2}}y_{u_{1\bar\eta}}\right)+\frac{1}{4}\ln\frac{m^2_{\tilde\nu_R}}{Q^2}y_{u_R}y_{\bar\eta_R}\Bigg], \tag{A11}$$

$$
\begin{aligned}
\Delta m^2_{\Phi_\eta\Phi_\eta} &= \frac{3}{16\pi^2}\left\{\frac{1}{1152}[f(Q^2,m^2_{\tilde{t}_1})-f(Q^2,m^2_{\tilde{t}_2})]\left[\frac{1}{m^2_{\tilde{t}_1}-m^2_{\tilde{t}_2}}-\frac{M^4_{tm}}{(m^2_{\tilde{t}_1}-m^2_{\tilde{t}_2})^3}\right]C_3^4u_1^2\right.\\
&\quad\left.+\frac{1}{9216}\ln\frac{m^2_{\tilde{t}_1}}{Q^2}\left(C_4^2u_1+\frac{M^2_{tm}C_3^2u_1}{m^2_{\tilde{t}_1}-m^2_{\tilde{t}_2}}\right)^2+\frac{1}{9216}\ln\frac{m^2_{\tilde{t}_2}}{Q^2}\left(C_4^2u_1-\frac{M^2_{tm}C_3^2u_1}{m^2_{\tilde{t}_1}-m^2_{\tilde{t}_2}}\right)^2\right\}\\
&\quad+\frac{3}{16\pi^2}\left\{\frac{1}{1152}[f(Q^2,m^2_{\tilde{b}_1})-f(Q^2,m^2_{\tilde{b}_2})]\left[\frac{1}{m^2_{\tilde{b}_1}-m^2_{\tilde{b}_2}}-\frac{M^4_{bm}}{(m^2_{\tilde{b}_1}-m^2_{\tilde{b}_2})^3}\right]C_3^4u_2^2\right.\\
&\quad\left.+\frac{1}{9216}\ln\frac{m^2_{\tilde{b}_1}}{Q^2}\left(C_4^2u_2+\frac{M^2_{bm}C_3^2u_2}{m^2_{\tilde{b}_1}-m^2_{\tilde{b}_2}}\right)^2+\frac{1}{9216}\ln\frac{m^2_{\tilde{b}_2}}{Q^2}\left(C_4^2u_2-\frac{M^2_{bm}C_3^2u_2}{m^2_{\tilde{b}_1}-m^2_{\tilde{b}_2}}\right)^2\right\}\\
&\quad+\frac{1}{32\pi^2}\left[f(Q^2,m^2_{\tilde{\nu}_R})\left(\frac{1}{2}g_{YB}g_B+4Y^2_{x,ij}\right)+f(Q,m^2_{\tilde{\nu}_R})\left(\frac{1}{8}Y_1^{-\frac{3}{2}}y_\eta^2-\frac{1}{4}Y_1^{-\frac{1}{2}}y_{\eta_1}\right)\right.\\
&\quad\left.+\frac{1}{4}\ln\frac{m^2_{\tilde{\nu}_R}}{Q^2}y^2_{\eta_R}-8f(Q^2,m^2_{\tilde{\nu}_R})Y^2_{x,ij}+8\ln\frac{m^2_{\tilde{\nu}_R}}{Q^2}u_1^2Y^4_{x,ij}\right],
\end{aligned}
\tag{A12}
$$

$$
\begin{aligned}
\Delta m^2_{\Phi_{\bar\eta}\Phi_\eta} &= \frac{3}{16\pi^2}\left\{\frac{1}{1152}[f(Q^2,m^2_{\tilde{t}_1})-f(Q^2,m^2_{\tilde{t}_2})]\left[-\frac{1}{m^2_{\tilde{t}_1}-m^2_{\tilde{t}_2}}+\frac{M^4_{tm}}{(m^2_{\tilde{t}_1}-m^2_{\tilde{t}_2})^3}\right]C_3^4u_2u_1\right.\\
&\quad+\frac{1}{9216}\ln\frac{m^2_{\tilde{t}_1}}{Q^2}\left(-C_4^2u_2-\frac{M^2_{tm}C_3^2u_1}{m^2_{\tilde{t}_1}-m^2_{\tilde{t}_2}}\right)\times\left(C_4^2u_1+\frac{M^2_{tm}C_3^2u_2}{m^2_{\tilde{t}_1}-m^2_{\tilde{t}_2}}\right)\\
&\quad\left.+\frac{1}{9216}\ln\frac{m^2_{\tilde{t}_2}}{Q^2}\left(-C_4^2u_2+\frac{M^2_{tm}C_3^2u_1}{m^2_{\tilde{t}_1}-m^2_{\tilde{t}_2}}\right)\times\left(C_4^2u_1-\frac{M^2_{tm}C_3^2u_2}{m^2_{\tilde{t}_1}-m^2_{\tilde{t}_2}}\right)\right\}\\
&\quad+\frac{3}{16\pi^2}\left\{\frac{1}{1152}[f(Q^2,m^2_{\tilde{b}_1})-f(Q^2,m^2_{\tilde{b}_2})]\left[-\frac{1}{m^2_{\tilde{b}_1}-m^2_{\tilde{b}_2}}+\frac{M^4_{bm}}{(m^2_{\tilde{b}_1}-m^2_{\tilde{b}_2})^3}\right]C_3^4u_2u_1\right.\\
&\quad+\frac{1}{9216}\ln\frac{m^2_{\tilde{b}_1}}{Q^2}\left(-C_4^2u_1-\frac{M^2_{bm}C_3^2u_2}{m^2_{\tilde{b}_1}-m^2_{\tilde{b}_2}}\right)\times\left(C_4^2u_2+\frac{M^2_{bm}C_3^2u_1}{m^2_{\tilde{b}_1}-m^2_{\tilde{b}_2}}\right)\\
&\quad\left.+\frac{1}{9216}\ln\frac{m^2_{\tilde{b}_2}}{Q^2}\left(-C_4^2u_1+\frac{M^2_{bm}C_3^2u_2}{m^2_{\tilde{b}_1}-m^2_{\tilde{b}_2}}\right)\times\left(C_4^2u_2-\frac{M^2_{bm}C_3^2u_1}{m^2_{\tilde{b}_1}-m^2_{\tilde{b}_2}}\right)\right\}\\
&\quad+\frac{1}{32\pi^2}\left[f(Q^2,m^2_{\tilde{\nu}_R})\left(\frac{1}{8}Y_1^{-\frac{3}{2}}y_\eta y_{\bar\eta}-\frac{1}{4}Y_1^{-\frac{1}{2}}y_{\eta_{1\bar\eta}}\right)+\frac{1}{4}\ln\frac{m^2_{\tilde{\nu}_R}}{Q^2}y_{\eta_R}y_{\bar\eta_R}\right],
\end{aligned}
\tag{A13}
$$

$$
\begin{aligned}
\Delta m^2_{\Phi_{\bar\eta}\Phi_{\bar\eta}} &= \frac{3}{16\pi^2}\left\{\frac{1}{1152}[f(Q^2,m^2_{\tilde{t}_1})-f(Q^2,m^2_{\tilde{t}_2})]\left[\frac{1}{m^2_{\tilde{t}_1}-m^2_{\tilde{t}_2}}-\frac{M^4_{tm}}{(m^2_{\tilde{t}_1}-m^2_{\tilde{t}_2})^3}\right]C_3^4u_2^2\right.\\
&\quad\left.+\frac{1}{192}\ln\frac{m^2_{\tilde{t}_1}}{Q^2}\left(-C_4^2u_2-\frac{M^2_{tm}C_3^2u_2}{m^2_{\tilde{t}_1}-m^2_{\tilde{t}_2}}\right)^2+\frac{1}{192}\ln\frac{m^2_{\tilde{t}_2}}{Q^2}\left(-C_4^2u_2+\frac{M^2_{tm}C_3^2u_2}{m^2_{\tilde{t}_1}-m^2_{\tilde{t}_2}}\right)^2\right\}\\
&\quad+\frac{3}{16\pi^2}\left\{\frac{1}{1152}[f(Q^2,m^2_{\tilde{b}_1})-f(Q^2,m^2_{\tilde{b}_2})]\left[\frac{1}{m^2_{\tilde{b}_1}-m^2_{\tilde{b}_2}}-\frac{M^4_{bm}}{(m^2_{\tilde{b}_1}-m^2_{\tilde{b}_2})^3}\right]C_3^4u_1^2\right.\\
&\quad\left.+\frac{1}{192}\ln\frac{m^2_{\tilde{b}_1}}{Q^2}\left(-C_4^2u_1-\frac{M^2_{bm}C_3^2u_1}{m^2_{\tilde{b}_1}-m^2_{\tilde{b}_2}}\right)^2+\frac{1}{192}\ln\frac{m^2_{\tilde{b}_2}}{Q^2}\left(-C_4^2u_1+\frac{M^2_{bm}C_3^2u_1}{m^2_{\tilde{b}_1}-m^2_{\tilde{b}_2}}\right)^2\right\}\\
&\quad+\frac{1}{32\pi^2}\left[-\frac{1}{2}f(Q^2,m^2_{\tilde{\nu}_R})g_{YB}g_B+f(Q,m^2_{\tilde{\nu}_R})\left(\frac{1}{8}Y_1^{-\frac{3}{2}}y^2_{\bar\eta}-\frac{1}{4}Y_1^{-\frac{1}{2}}y_{\bar\eta_1}\right)+\frac{1}{4}\ln\frac{m^2_{\tilde{\nu}_R}}{Q^2}y^2_{\bar\eta_R}\right].
\end{aligned}
\tag{A14}
$$

Here, $f(Q^2,m^2_{t,\tilde{t}_{1,2}})=\frac{1}{4}m^2_{t,\tilde{t}_{1,2}}(\ln\frac{m^2_{t,\tilde{t}_{1,2}}}{Q^2}-1)$, $f(Q^2,m^2_{b,\tilde{b}_{1,2}})=\frac{1}{4}m^2_{b,\tilde{b}_{1,2}}(\ln\frac{m^2_{b,\tilde{b}_{1,2}}}{Q^2}-1)$, $f(Q^2,m^2_{\nu_{iR},\tilde{\nu}_{iR}})=\frac{1}{4}m^2_{\nu_{iR},\tilde{\nu}_{iR}}(\ln\frac{m^2_{\nu_{iR},\tilde{\nu}_{iR}}}{Q^2}-1)$, and

$$C_1^2=-6g_2^2+10(g_1^2+g_{YB}^2)+4g_{YB}g_B,\quad C_2^2=-6g_2^2-6(g_1^2+g_{YB}^2)+24Y_t^2,\quad C_2'^2=-6g_2^2-6(g_1^2+g_{YB}^2)+24Y_b^2,$$
$$C_3^2=-8g_B^2-20g_{YB}g_B,\quad C_4^2=12g_{YB}g_B,\quad F_2=(v_2A_t-v_1\mu)A_tY_t^2,\quad F_2'=(v_1A_b-v_2\mu)A_bY_b^2,$$
$$F_3=-A_tY_t^2\mu,\quad F_4=(v_1\mu-v_2A_t)\mu Y_t^2,\quad F_4'=(v_2\mu-v_1A_b)\mu Y_b^2,\quad F_5=\mu^2Y_t^2,\quad F_5'=\mu^2Y_b^2,$$
$$Y_1=(m^2_{\tilde{\nu}^I_{LL}}+m^2_{\tilde{\nu}^I_{RR}})^2-4(m^2_{\tilde{\nu}^I_{LL}}m^2_{\tilde{\nu}^I_{RR}}-m^4_{\tilde{\nu}^I_{LR}}),\quad y_d=\frac{1}{4}(m^2_{\tilde{\nu}^I_{LL}}+m^2_{\tilde{\nu}^I_{RR}})g^2v_1-(g^2+g_{YB}g_B)v_1m^2_{\tilde{\nu}^I_{RR}}+\frac{1}{4}m^2_{\tilde{\nu}^I_{LL}}g_{YB}g_B(v_1-\sqrt{2}\mu Y_{\nu,ij}),$$
$$y_u=\left(-\frac{1}{2}g^2v_2+4v_2Y^2_{\nu,ij}\right)(m^2_{\tilde{\nu}^I_{LL}}+m^2_{\tilde{\nu}^I_{RR}})-4\left[-\frac{1}{4}(g^2+g_{YB}g_B)v_2m^2_{\tilde{\nu}^I_{RR}}+v_2Y^2_{\nu,ij}m^2_{\tilde{\nu}^I_{RR}}\right.$$
$$\left.+\left(\frac{1}{4}g_{YB}g_B+Y^2_{\nu,ij}\right)m^2_{\tilde{\nu}^I_{LL}}v_2\right]+(\sqrt{2}T^{ij}_{\nu}+2u_1Y_{x,ij}Y_{\nu,ij})m^2_{\tilde{\nu}^I_{LR}},$$
$$y_\eta=(g_{YB}g_Bu_1+2\sqrt{2}T^{ij}_x+8u_1Y^2_{x,ij})(m^2_{\tilde{\nu}^I_{RR}}+m^2_{\tilde{\nu}^I_{LL}})-2(g_B^2+g_{YB}g_B)u_1m^2_{\tilde{\nu}^I_{RR}}$$
$$-m^2_{\tilde{\nu}^I_{LL}}(-2g_B^2u_1+4\sqrt{2}T^{ij}_x+16u_1Y^2_{x,ij})+2v_2Y_{x,ij}Y_{\nu,ij}m^2_{\tilde{\nu}^I_{LR}},$$
$$y_{\bar{\eta}}=[g_B^2u_1-\sqrt{2}\mu Y_{x,ij}-(g_B^2+g_{YB}g_B)u_2](m^2_{\tilde{\nu}^I_{RR}}+m^2_{\tilde{\nu}^I_{LL}})-2(g_B^2+g_{YB}g_B)u_2m^2_{\tilde{\nu}^I_{RR}}+m^2_{\tilde{\nu}^I_{LL}}(2g_B^2u_2-4\sqrt{2}\mu Y_{x,ij}),$$
$$y_{d_1}=\frac{1}{2}g^2(m^2_{\tilde{\nu}^I_{LL}}+m^2_{\tilde{\nu}^I_{RR}})+\frac{1}{8}g^4v_1^2-(g^2+g_{YB}g_B)m^2_{\tilde{\nu}^I_{RR}}+\frac{1}{2}g_{YB}g_B(g^2+g_{YB}g_B)v_1^2+g_{YB}g_Bm^2_{\tilde{\nu}^I_{LL}}+\mu^2Y^2_{\nu,ij},$$
$$y_{u_1}=\left(4Y^2_{\nu,ij}-\frac{1}{2}g^2\right)(m^2_{\tilde{\nu}^I_{LL}}+m^2_{\tilde{\nu}^I_{RR}})+\left(4Y^2_{\nu,ij}-\frac{1}{2}g^2\right)\left(2Y^2_{\nu,ij}-\frac{1}{4}g^2\right)v_2^2$$
$$-4\left[-\frac{1}{4}(g^2+g_{YB}g_B+Y^2_{\nu,ij})m^2_{\tilde{\nu}^I_{RR}}+m^2_{\tilde{\nu}^I_{LL}}\left(\frac{1}{4}g_{YB}g_B+Y^2_{\nu,ij}\right)-\frac{1}{2}(g^2+g_{YB}g_B+Y^2_{\nu,ij})\left(\frac{1}{4}g_{YB}g_B+Y^2_{\nu,ij}\right)v_2^2\right]$$
$$+\left(\frac{\sqrt{2}}{4}T^{ij}_{\nu}+\frac{1}{2}u_1Y_{x,ij}Y_{\nu,ij}\right)^2,$$
$$y_{\eta_1}=(g_{YB}g_B+8Y^2_{x,ij})(m^2_{\tilde{\nu}^I_{LL}}+m^2_{\tilde{\nu}^I_{RR}})-2(g_B^2+g_{YB}g_B)m^2_{\tilde{\nu}^I_{RR}}-u_1(g_B^2+g_{YB}g_B)\left(-\frac{1}{4}g_B^2u_1+\frac{\sqrt{2}}{2}T^{ij}_x+8u_1Y^2_{x,ij}\right)$$
$$-(g_B^2+g_{YB}g_B)\left(-g_B^2u_1+\frac{\sqrt{2}}{2}T^{ij}_x+8u_1Y^2_{x,ij}\right)u_1-m^2_{\tilde{\nu}^I_{LL}}(-2g_B^2+16Y^2_{x,ij})+(v_2Y_{x,ij}Y_{\nu,ij})^2,$$
$$y_{\bar{\eta}_1}=-g_{YB}g_B(m^2_{\tilde{\nu}^I_{LL}}+m^2_{\tilde{\nu}^I_{RR}})+2(g_B^2+g_{YB}g_B)m^2_{\tilde{\nu}^I_{RR}}+\frac{1}{2}g_B^2m^2_{\tilde{\nu}^I_{LL}}+u_2(g_B^2+g_{YB}g_B)(g_B^2u_2-2\sqrt{2}\mu Y_{x,ij})$$
$$+(-g_B^2u_2+2\sqrt{2}\mu Y_{x,ij})(g_B^2+g_{YB}g_B)u_2,$$
$$y_{u_{1\eta}}=\left(-\frac{1}{2}g^2v_2+4v_2Y^2_{\nu,ij}\right)\left(\frac{1}{2}g_{YB}g_Bu_1+\sqrt{2}T^{ij}_x+4u_1Y^2_{x,ij}\right)-4(g^2+g_{YB}g_B+Y^2_{\nu,ij})\left(\frac{1}{8}g_B^2u_1-\frac{\sqrt{2}}{4}T^{ij}_x-u_1Y^2_{x,ij}\right)v_2$$
$$-2(g_B^2+g_{YB}g_B)\left(\frac{1}{4}g_{YB}g_B+Y^2_{\nu,ij}\right)v_2u_1+\left(2u_1Y_{x,ij}Y_{\nu,ij}-\frac{\sqrt{2}}{2}\mu Y_{\nu,ij}\right)v_2,$$
$$y_{\eta_{1\bar{\eta}}}=-2(g_B^2+g_{YB}g_B)\left(\frac{1}{2}g_B^2u_2-\sqrt{2}\mu Y_{x,ij}\right)u_1-(g_B^2+g_{YB}g_B)(-g_B^2u_1+2\sqrt{2}T^{ij}_x+8u_1Y^2_{x,ij})u_2,$$
$$y_{d_R}=\frac{1}{8}g^2v_1-\frac{1}{4}Y_1^{-\frac{1}{2}}y_d,\quad y_{u_R}=-\frac{1}{8}g^2v_2-\frac{1}{4}Y_1^{-\frac{1}{2}}y_u,\quad y_{\eta_R}=\frac{1}{4}g_{YB}g_Bu_1+\frac{\sqrt{2}}{2}T^{ij}_x+2u_1Y^2_{x,ij}-\frac{1}{4}Y_1^{-\frac{1}{2}}y_\eta,$$
$$y_{\bar{\eta}_R}=-\frac{1}{4}g_{YB}g_Bu_2-\sqrt{2}\mu Y_{x,ij}-\frac{1}{4}Y_1^{-\frac{1}{2}}y_{\bar{\eta}}. \tag{A15}$$

In addition,

$$m^2_{\tilde{t}_{1,2}}=\frac{1}{2}M^2_{ts}\pm\sqrt{M^4_{tm}+2M^2_{tLR}}, \tag{A16}$$

$$M^2_{ts}=\frac{1}{8}(g^2v^2\cos 2\beta+4g_{YB}g_Bu^2\cos 2\beta')+m^2_{\tilde{q},33}+\frac{1}{2}v_2^2Y_t^2, \tag{A17}$$

$$M_{tm}^2 = \frac{1}{24}\{[3g_2^2 - 5(g_1^2 + g_{YB}^2) - 2g_{YB}g_B]v^2\cos 2\beta + (8g_B^2 - 20g_{YB}g_B)u^2\cos 2\beta' + m_{\tilde{u},33}^2 + \frac{1}{2}v_2^2Y_t^2\}, \tag{A18}$$

$$M_{tLR}^2 = \frac{1}{2}(-v_1\mu Y_t + v_2A_tY_t)^2, \tag{A19}$$

$$m_{\tilde{b}_{1,2}}^2 = \frac{1}{2}M_{bs}^2 \pm \sqrt{M_{bm}^4 + 2M_{bLR}^2}, \tag{A20}$$

$$M_{bs}^2 = \frac{1}{8}(g^2v^2\cos 2\beta + 4g_{YB}g_Bu^2\cos 2\beta') + m_{\tilde{q},33}^2 + \frac{1}{2}v_1^2Y_b^2, \tag{A21}$$

$$M_{bm}^2 = \frac{1}{24}\{[3g_2^2 - 5(g_1^2 + g_{YB}^2) - 2g_{YB}g_B]v^2\cos 2\beta + (8g_B^2 - 20g_{YB}g_B)u^2\cos 2\beta' + m_{\tilde{d},33}^2 + \frac{1}{2}v_1^2Y_b^2\}, \tag{A22}$$

$$M_{bLR}^2 = \frac{1}{2}(-v_2\mu Y_b + v_1A_bY_b)^2, \tag{A23}$$

and

$$m_{\nu_R}^2 = 2u_1^2Y_{x,ij}^2. \tag{A24}$$

$$m_{\tilde{\nu}_R}^2 = \frac{1}{2}(m_{\tilde{\nu}_{LL}^I}^2 + m_{\tilde{\nu}_{RR}^I}^2) - \frac{1}{2}\sqrt{(m_{\tilde{\nu}_{LL}^I}^2 + m_{\tilde{\nu}_{RR}^I}^2)^2 - 4(m_{\tilde{\nu}_{LL}^I}^2 m_{\tilde{\nu}_{RR}^I}^2 - m_{\tilde{\nu}_{LR}^I}^4)}, \tag{A25}$$

$$m_{\tilde{\nu}_{LL}^I}^2 = \frac{1}{4}(g_B^2 + g_{YB}g_B)u^2\cos 2\beta' + \frac{1}{8}(g^2 + g_{YB}g_B)v^2\cos 2\beta + m_{\tilde{L},ij}^2 + \frac{1}{2}v_2^2Y_{\nu,ij}^2, \tag{A26}$$

$$m_{\tilde{\nu}_{RR}^I}^2 = -\frac{1}{4}g_B^2u^2\cos 2\beta' - \frac{1}{8}g_{YB}g_Bv^2\cos 2\beta + m_{\tilde{\nu},ij}^2 - \sqrt{2}u_2\mu Y_{x,ij} + \frac{1}{2}v_2^2Y_{\nu,ij}^2 + \sqrt{2}u_1T_x^{ij}, \tag{A27}$$

$$m_{\tilde{\nu}_{LR}^I}^2 = \frac{1}{\sqrt{2}}v_2T_\nu^{ij} - \frac{1}{\sqrt{2}}\mu v_1Y_{\nu,ij} + u_1v_2Y_{\nu,ij}Y_{x,ij}. \tag{A28}$$

## APPENDIX B: THE TREE LEVEL CORRECTION OF HIGGS SELF-COUPLING

$$\begin{aligned}
\lambda_{h_ih_jh_k}^{(0)} = \frac{1}{4}&(Z_{i1}^H((g_1^2 + g_{YB}^2 + g_2^2)Z_{j2}^H(v_1Z_{k2}^H + v_2Z_{k1}^H) - 2g_{YB}g_B(Z_{j3}^H(v_1Z_{k3}^H + u_1Z_{k1}^H) - Z_{j4}^H(u_2Z_{k1}^H + v_1Z_{k4}^H))\\
&+ Z_{j1}^H(-2g_{YB}g_B(-u_2Z_{k4}^H + u_1Z_{k3}^H) - 3(g_1^2 + g_{YB}^2 + g_2^2)v_1Z_{k1}^H + (g_1^2 + g_{YB}^2 + g_2^2)v_2Z_{k2}^H))\\
&+ Z_{i2}^H((g_1^2 + g_{YB}^2 + g_2^2)Z_{j1}^H(v_1Z_{k2}^H + v_2Z_{k1}^H) + 2g_{YB}g_B(Z_{j3}^H(u_1Z_{k2}^H + v_2Z_{k3}^H) - Z_{j4}^H(u_2Z_{k2}^H + v_2Z_{k4}^H))\\
&+ Z_{j2}^H(2g_{YB}g_B)(-u_2Z_{k4}^H + u_1Z_{k3}^H) - 3(g_1^2 + g_{YB}^2 + g_2^2)v_2Z_{k2}^H + (g_1^2 + g_{YB}^2 + g_2^2)v_1Z_{k1}^H))\\
&- 2(-Z_{i4}^H(-g_{YB}g_Bu_2Z_{j2}^HZ_{k2}^H + 2g_B^2u_2Z_{j3}^HZ_{k3}^H - g_{YB}g_Bv_2Z_{j2}^HZ_{k4}^H + 2g_B^2u_1Z_{j3}^HZ_{k4}^H + g_{YB}g_BZ_{j1}^H(u_2Z_{k1}^H + v_1Z_{k4}^H)\\
&+ 2Z_{j4}^Hg_B^2(-3u_2Z_{k4}^H + u_1Z_{k3}^H) + g_{YB}g_Bv_1Z_{k1}^H - g_{YB}g_Bv_2Z_{k2}^H))\\
&+ Z_{i3}^H(-g_{YB}g_Bu_1Z_{j2}^HZ_{k2}^H - g_{YB}g_Bv_2Z_{j2}^HZ_{k3}^H - 2g_B^2u_2Z_{j4}^HZ_{k3}^H + g_{YB}g_BZ_{j1}^H(v_1Z_{k3}^H + u_1Z_{k1}^H)\\
&- 2g_B^2u_1Z_{j4}^HZ_{k4}^H + Z_{j3}^H(2g_B^2(3u_1Z_{k3}^H - u_2Z_{k4}^H) + g_{YB}g_Bv_1Z_{k1}^H - g_{YB}g_Bv_2Z_{k2}^H)))).
\end{aligned} \tag{B1}$$

## APPENDIX C: THE COUPLING OF TWO CP-ODD HIGGS AND ONE CP-EVEN HIGGS

$$
\begin{aligned}
\lambda_{A_iA_jh_k} = {} & \frac{1}{2!}\frac{\partial^3 V}{\partial(\mathrm{ImH}_1^1)^2\partial\mathrm{ReH}_1^1}Z^A_{1i}Z^A_{1j}Z^H_{1k} + \frac{1}{2!}\frac{\partial^3 V}{\partial(\mathrm{ImH}_2^2)^2\partial\mathrm{ReH}_1^1}Z^A_{2i}Z^A_{2j}Z^H_{1k} + \frac{1}{2!}\frac{\partial^3 V}{\partial(\partial\mathrm{Im}\tilde{\eta}_1)^2\partial\mathrm{ReH}_1^1}Z^A_{3i}Z^A_{3j}Z^H_{1k} \\
& + \frac{1}{2!}\frac{\partial^3 V}{\partial(\partial\mathrm{Im}\tilde{\eta}_2)^2\partial\mathrm{ReH}_1^1}Z^A_{4i}Z^A_{4j}Z^H_{1k} + \frac{\partial^3 V}{\partial\mathrm{ImH}_1^1\partial\mathrm{ImH}_2^2\partial\mathrm{ReH}_1^1}Z^A_{1i}Z^A_{2j}Z^H_{1k} + \frac{\partial^3 V}{\partial\mathrm{ImH}_1^1\partial\partial\mathrm{Im}\tilde{\eta}_1\partial\mathrm{ReH}_1^1}Z^A_{1i}Z^A_{3j}Z^H_{1k} \\
& + \frac{\partial^3 V}{\partial\mathrm{ImH}_1^1\partial\partial\mathrm{Im}\tilde{\eta}_2\partial\mathrm{ReH}_1^1}Z^A_{1i}Z^A_{4j}Z^H_{1k} + \frac{\partial^3 V}{\partial\mathrm{ImH}_2^2\partial\partial\mathrm{Im}\tilde{\eta}_1\partial\mathrm{ReH}_1^1}Z^A_{2i}Z^A_{3j}Z^H_{1k} + \frac{\partial^3 V}{\partial\mathrm{ImH}_2^2\partial\partial\mathrm{Im}\tilde{\eta}_2\partial\mathrm{ReH}_1^1}Z^A_{2i}Z^A_{4j}Z^H_{1k} \\
& + \frac{\partial^3 V}{\partial\mathrm{Im}\tilde{\eta}_1\partial\partial\mathrm{Im}\tilde{\eta}_2\partial\mathrm{ReH}_1^1}Z^A_{3i}Z^A_{4j}Z^H_{1k} + \frac{1}{2!}\frac{\partial^3 V}{\partial(\mathrm{ImH}_1^1)^2\partial\mathrm{ReH}_2^2}Z^A_{1i}Z^A_{1j}Z^H_{2k} + \frac{1}{2!}\frac{\partial^3 V}{\partial(\mathrm{ImH}_2^2)^2\partial\mathrm{ReH}_2^2}Z^A_{2i}Z^A_{2j}Z^H_{2k} \\
& + \frac{1}{2!}\frac{\partial^3 V}{\partial(\partial\mathrm{Im}\tilde{\eta}_1)^2\partial\mathrm{ReH}_2^2}Z^A_{3i}Z^A_{3j}Z^H_{2k} + \frac{1}{2!}\frac{\partial^3 V}{\partial(\partial\mathrm{Im}\tilde{\eta}_2)^2\partial\mathrm{ReH}_2^2}Z^A_{4i}Z^A_{4j}Z^H_{2k} + \frac{\partial^3 V}{\partial\mathrm{ImH}_1^1\partial\mathrm{ImH}_2^2\partial\mathrm{ReH}_2^2}Z^A_{1i}Z^A_{2j}Z^H_{2k} \\
& + \frac{\partial^3 V}{\partial\mathrm{ImH}_1^1\partial\partial\mathrm{Im}\tilde{\eta}_1\partial\mathrm{ReH}_2^2}Z^A_{1i}Z^A_{3j}Z^H_{2k} + \frac{\partial^3 V}{\partial\mathrm{ImH}_1^1\partial\partial\mathrm{Im}\tilde{\eta}_2\partial\mathrm{ReH}_2^2}Z^A_{1i}Z^A_{4j}Z^H_{2k} + \frac{\partial^3 V}{\partial\mathrm{ImH}_2^2\partial\partial\mathrm{Im}\tilde{\eta}_1\partial\mathrm{ReH}_2^2}Z^A_{2i}Z^A_{3j}Z^H_{2k} \\
& + \frac{\partial^3 V}{\partial\mathrm{ImH}_2^2\partial\partial\mathrm{Im}\tilde{\eta}_2\partial\mathrm{ReH}_2^2}Z^A_{2i}Z^A_{4j}Z^H_{2k} + \frac{\partial^3 V}{\partial\mathrm{Im}\tilde{\eta}_1\partial\partial\mathrm{Im}\tilde{\eta}_2\partial\mathrm{ReH}_2^2}Z^A_{3i}Z^A_{4j}Z^H_{2k} + \frac{1}{2!}\frac{\partial^3 V}{\partial(\mathrm{ImH}_1^1)^2\partial\mathrm{Re}\tilde{\eta}_1}Z^A_{1i}Z^A_{1j}Z^H_{3k} \\
& + \frac{1}{2!}\frac{\partial^3 V}{\partial(\mathrm{ImH}_2^2)^2\partial\mathrm{Re}\tilde{\eta}_1}Z^A_{2i}Z^A_{2j}Z^H_{3k} + \frac{1}{2!}\frac{\partial^3 V}{\partial(\partial\mathrm{Im}\tilde{\eta}_1)^2\partial\mathrm{Re}\tilde{\eta}_1}Z^A_{3i}Z^A_{3j}Z^H_{3k} + \frac{1}{2!}\frac{\partial^3 V}{\partial(\partial\mathrm{Im}\tilde{\eta}_2)^2\partial\mathrm{Re}\tilde{\eta}_1}Z^A_{4i}Z^A_{4j}Z^H_{3k} \\
& + \frac{\partial^3 V}{\partial\mathrm{ImH}_1^1\partial\mathrm{ImH}_2^2\partial\mathrm{Re}\tilde{\eta}_1}Z^A_{1i}Z^A_{2j}Z^H_{3k} + \frac{\partial^3 V}{\partial\mathrm{ImH}_1^1\partial\partial\mathrm{Im}\tilde{\eta}_1\partial\mathrm{Re}\tilde{\eta}_1}Z^A_{1i}Z^A_{3j}Z^H_{3k} + \frac{\partial^3 V}{\partial\mathrm{ImH}_1^1\partial\partial\mathrm{Im}\tilde{\eta}_2\partial\mathrm{Re}\tilde{\eta}_1}Z^A_{1i}Z^A_{4j}Z^H_{3k} \\
& + \frac{\partial^3 V}{\partial\mathrm{ImH}_2^2\partial\partial\mathrm{Im}\tilde{\eta}_1\partial\mathrm{Re}\tilde{\eta}_1}Z^A_{2i}Z^A_{3j}Z^H_{3k} + \frac{\partial^3 V}{\partial\mathrm{ImH}_2^2\partial\partial\mathrm{Im}\tilde{\eta}_2\partial\mathrm{Re}\tilde{\eta}_1}Z^A_{2i}Z^A_{4j}Z^H_{3k} + \frac{\partial^3 V}{\partial\mathrm{Im}\tilde{\eta}_1\partial\partial\mathrm{Im}\tilde{\eta}_2\partial\mathrm{Re}\tilde{\eta}_1}Z^A_{3i}Z^A_{4j}Z^H_{3k} \\
& + \frac{1}{2!}\frac{\partial^3 V}{\partial(\mathrm{ImH}_1^1)^2\partial\mathrm{Re}\tilde{\eta}_2}Z^A_{1i}Z^A_{1j}Z^H_{4k} + \frac{1}{2!}\frac{\partial^3 V}{\partial(\mathrm{ImH}_2^2)^2\partial\mathrm{Re}\tilde{\eta}_2}Z^A_{2i}Z^A_{2j}Z^H_{4k} + \frac{1}{2!}\frac{\partial^3 V}{\partial(\partial\mathrm{Im}\tilde{\eta}_1)^2\partial\mathrm{Re}\tilde{\eta}_2}Z^A_{3i}Z^A_{3j}Z^H_{4k} \\
& + \frac{1}{2!}\frac{\partial^3 V}{\partial(\partial\mathrm{Im}\tilde{\eta}_2)^2\partial\mathrm{Re}\tilde{\eta}_2}Z^A_{4i}Z^A_{4j}Z^H_{4k} + \frac{\partial^3 V}{\partial\mathrm{ImH}_1^1\partial\mathrm{ImH}_2^2\partial\mathrm{Re}\tilde{\eta}_2}Z^A_{1i}Z^A_{2j}Z^H_{4k} + \frac{\partial^3 V}{\partial\mathrm{ImH}_1^1\partial\partial\mathrm{Im}\tilde{\eta}_1\partial\mathrm{Re}\tilde{\eta}_2}Z^A_{1i}Z^A_{3j}Z^H_{4k} \\
& + \frac{\partial^3 V}{\partial\mathrm{ImH}_1^1\partial\partial\mathrm{Im}\tilde{\eta}_2\partial\mathrm{Re}\tilde{\eta}_2}Z^A_{1i}Z^A_{4j}Z^H_{4k} + \frac{\partial^3 V}{\partial\mathrm{ImH}_2^2\partial\partial\mathrm{Im}\tilde{\eta}_1\partial\mathrm{Re}\tilde{\eta}_2}Z^A_{2i}Z^A_{3j}Z^H_{4k} \\
& + \frac{\partial^3 V}{\partial\mathrm{ImH}_2^2\partial\partial\mathrm{Im}\tilde{\eta}_2\partial\mathrm{Re}\tilde{\eta}_2}Z^A_{2i}Z^A_{4j}Z^H_{4k} + \frac{\partial^3 V}{\partial\mathrm{Im}\tilde{\eta}_1\partial\partial\mathrm{Im}\tilde{\eta}_2\partial\mathrm{Re}\tilde{\eta}_2}Z^A_{3i}Z^A_{4j}Z^H_{4k}. \qquad \text{(C1)}
\end{aligned}
$$

Here, the detailed expression about tree level correction $\lambda^{(0)}_{A_iA_jh_k}$ is written as:

$$
\begin{aligned}
\lambda^{(0)}_{A_iA_jh_k} = {} & \frac{1}{4}(Z^A_{i1}Z^A_{j1}(-2g_{YB}g_B(-u_2Z^H_{k4} + u_1Z^H_{k3}) - (g_1^2 + g_{YB}^2 + g_2^2)v_1Z^H_{k1} + (g_1^2 + g_{YB}^2 + g_2^2)v_2Z^H_{k2}) \\
& + Z^A_{i2}Z^A_{j2}(2g_{YB}g_B(-u_2Z^H_{k4} + u_1Z^H_{k3}) + (g_1^2 + g_{YB}^2 + g_2^2)v_1Z^H_{k1} - (g_1^2 + g_{YB}^2 + g_2^2)v_2Z^H_{k2}) \\
& - 2(Z^A_{i3}Z^A_{j3} - Z^A_{i4}Z^A_{j4})(2g_B^2(-u_2Z^H_{k4} + u_1Z^H_{k3}) + g_{YB}g_Bv_1Z^H_{k1} - g_{YB}g_Bv_2Z^H_{k2})). \qquad \text{(C2)}
\end{aligned}
$$

## APPENDIX D: THE WILSON COEFFICIENTS OF THE PROCESS $e^+e^- \to hhZ$

$C^{(1)}_{L,R}$ denotes the Wilson coefficient corresponding to Fig. 1:

$$C^{(1)}_{L,R} = C^a_{1L,R} + C^b_{1L,R} + C^c_{1L,R}. \qquad \text{(D1)}$$

$$
\begin{aligned}
C^a_{1L,R} &= \frac{i\lambda_{h_ih_jh_k}g_{NZh}g_{eeN1,2}}{[(p_1+p_2)^2 - m_N^2][(q_1+q_2)^2 - m_{h_i}^2]}, \\
C^b_{1L,R} &= \frac{ig_{NZh}g_{NNh}g_{eeN1,2}}{[(p_1+p_2)^2 - m_N^2][(p_1+p_2-q_1)^2 - m_N^2]}, \\
C^c_{1L,R} &= \frac{-ig_{ZZhh}g_{eeZ1,2}}{(p_1+p_2)^2 - m_Z^2}. \qquad \text{(D2)}
\end{aligned}
$$

Effective operator:

$$\mathcal{O}_{1L,R} = \bar{v}(p_1)(\gamma^\alpha)p_{L,R}u(p_2)\varepsilon^*_\alpha(k). \tag{D3}$$

$C^{(2)}_{L,R}$ represents the Wilson coefficient corresponding to Fig. 2:

$$C^{(2)}_{L,R} = C^{a'}_{1L,R} + C^{b'}_{2L,R} + C^{c'}_{3L,R} + C^{d'}_{4L,R}. \tag{D4}$$

$$\begin{aligned}
C^{a'}_{1L,R} &= \frac{-ig_{ZhA}g_{NhA}g_{eeN1,2}}{2[(p_1+p_2)^2-m_N^2][(p_1+p_2-q_1)^2-m_{A_k}^2]},\\
C^{b'}_{2L,R} &= \frac{2i\lambda_{h_ih_jh_k}g_{ZhA}g_{eeA1,2}}{[(p_1+p_2)^2-m_{A_k}^2][(p_1+p_2-k)^2-m_{h_i}^2]},\\
C^{c'}_{3L,R} &= \frac{-2i\lambda_{A_iA_jh_k}g_{eeA1,2}g_{ZhA}}{[(p_1+p_2)^2-m_{A_k}^2][(p_1+p_2-q_2)^2-m_{A_i}^2]},\\
C^{d'}_{4L,R} &= \frac{-ig_{NZh}g_{eeA1,2}g_{NhA}}{[(p_1+p_2)^2-m_{A_k}^2][(p_1+p_2-q_1)^2-m_N^2]}.
\end{aligned} \tag{D5}$$

Effective operator:

$$\begin{aligned}
\mathcal{O}_{2L,R} &= \bar{v}(p_1)(p_{1\alpha}+p_{2\alpha})p_{L,R}u(p_2)\varepsilon^*_\alpha(k),\\
\mathcal{O}_{3L,R} &= \bar{v}(p_1)(q_{2\alpha}-p_{1\alpha}-p_{2\alpha}-q_{1\alpha})p_{L,R}u(p_2)\varepsilon^*_\alpha(k);
\end{aligned} \tag{D6}$$

$\mathcal{O}_{1L,R}$, $\mathcal{O}_{4L,R}$ have been given in Eq. (D3) and Eq. (27), respectively. Furthermore, $\lambda_{h_ih_jh_k}$ has been given in Eq. (26) and Appendix B, and $\lambda_{A_iA_jh_k}$ denotes the triple Higgs self-coupling of $CP$-even and $CP$-odd Higgs; the detailed expression about this has been given in Appendix C. The superscripts (a, b, c, $a'$, $b'$, $c'$, $d'$) respectively represent the corresponding Feynman diagram labels in Figs. 1 and 2, and $m_h$, $m_A$ denote the masses for Higgs and pseudoscalar Higgs, with $i$, $k = 1, 2, 3, 4$ denoting the index of generation.

$$\begin{aligned}
g_{ZZhh} &= \frac{1}{2}g_2^2\cos\theta_W^2\cos\theta_W'^2 Z^H_{i1}Z^H_{j1} + g_1g_2\cos\theta_W\cos\theta_W'^2\sin\theta_W Z^H_{i1}Z^H_{j1} + \frac{1}{2}g_1^2\cos\theta_W'^2\sin\theta_W^2 Z^H_{i1}Z^H_{j1}\\
&\quad - g_{YB}g_2\cos\theta_W\cos\theta'_W\sin\theta'_W Z^H_{i1}Z^H_{j1} - g_1g_{YB}\cos\theta'_W\sin\theta_W\sin\theta'_W Z^H_{i1}Z^H_{j1} + \frac{1}{2}g_{YB}^2\sin\theta_W'^2 Z^H_{i1}Z^H_{j1} + \frac{1}{2}g_2^2\cos\theta_W^2\cos\theta_W'^2 Z^H_{i2}Z^H_{j2}\\
&\quad + g_1g_2\cos\theta_W\cos\theta_W'^2\sin\theta_W Z^H_{i2}Z^H_{j2} + \frac{1}{2}g_1^2\cos\theta_W'^2\sin\theta_W^2 Z^H_{i2}Z^H_{j2} - g_{YB}g_2\cos\theta_W\cos\theta'_W\sin\theta'_W Z^H_{i2}Z^H_{j2}\\
&\quad - g_1g_{YB}\cos\theta'_W\sin\theta_W\sin\theta'_W Z^H_{i2}Z^H_{j2} + \frac{1}{2}g_{YB}^2\sin\theta_W'^2 Z^H_{i2}Z^H_{j2} + +2g_B^2\sin\theta_W'^2 Z^H_{i3}Z^H_{j3} + 2g_B^2\sin\theta_W'^2 Z^H_{i4}Z^H_{j4}.
\end{aligned} \tag{D7}$$

$$\begin{aligned}
g_{eeZ1} &= -\frac{1}{2}g_1\cos\theta'_W\sin\theta_W + \frac{1}{2}g_2\cos\theta_W\cos\theta'_W + \frac{1}{2}(g_{YB}+g_B)\sin\theta'_W,\\
g_{eeZ2} &= -g_1\cos\theta'_W\sin\theta_W + \frac{1}{2}(2g_{YB}+g_B)\sin\theta'_W,\\
g_{eeZ'1} &= \frac{1}{2}((g_1\sin\theta_W - g_2\cos\theta_W)\sin\theta'_W + (g_{YB}+g_B)\cos\theta'_W),\\
g_{eeZ'2} &= g_1\sin\theta'_W\sin\theta_W - \frac{1}{2}(2g_{YB}+g_B)\cos\theta'_W.
\end{aligned} \tag{D8}$$

$$\begin{aligned}
g_{ZZh} &= \frac{1}{2}(v_1(g_1\cos\theta'_W\sin\theta_W + g_2\cos\theta_W\cos\theta'_W - g_{YB}\sin\theta'_W)^2 Z^H_{i1}\\
&\quad + v_2(g_1\cos\theta'_W\sin\theta_W + g_2\cos\theta_W\cos\theta'_W - g_{YB}\sin\theta'_W)^2 Z^H_{i2} + 4g_B^2\sin^2\theta'_W(u_2Z^H_{i4}+u_1Z^H_{i3})),\\
g_{Z'Z'h} &= \frac{1}{2}(v_1((g_1\sin\theta_W + g_2\cos\theta_W)\sin\theta'_W + g_{YB}\cos\theta'_W)^2 Z^H_{i1}\\
&\quad + v_2((g_1\sin\theta_W + g_2\cos\theta_W)\sin\theta'_W + g_{YB}\cos\theta'_W)^2 Z^H_{i2} + 4g_B^2\cos^2\theta'_W(u_2Z^H_{i4}+u_1Z^H_{i3})),\\
g_{ZZ'h} &= \frac{1}{2}(-v_1(g_1g_{YB}\cos\theta_W'^2\sin\theta_W + g_2^2\cos\theta_W^2\cos\theta'_W\sin\theta'_W + \cos\theta'_W(g_1^2\sin\theta_W^2 - g_{YB}^2)\sin\theta'_W - g_1g_{YB}\sin\theta_W\sin\theta_W'^2\\
&\quad + g_2\cos\theta_W(g_1\sin\theta_W\sin 2\theta'_W + g_{YB}\cos\theta_W'^2 - g_{YB}\sin\theta_W'^2))Z^H_{i1} - v_2(g_1g_{YB}\cos\theta_W'^2\sin\theta_W + g_2^2\cos\theta_W^2\cos\theta'_W\sin\theta'_W\\
&\quad + \cos\theta'_W(g_1^2\sin\theta_W^2 - g_{YB}^2)\sin\theta'_W - g_1g_{YB}\sin\theta_W\sin\theta_W'^2 + g_2\cos\theta_W(g_1\sin\theta_W\sin 2\theta'_W + g_{YB}\cos\theta_W'^2 - g_{YB}\sin\theta_W'^2))Z^H_{i2}\\
&\quad + 2g_B^2\sin 2\theta'_W(u_2Z^H_{i4}+u_1Z^H_{i3})).
\end{aligned} \tag{D9}$$

$$g_{ZAh} = \frac{1}{2}((g_1 \cos\theta'_W \sin\theta_W + g_2 \cos\theta_W \cos\theta'_W - g_{YB} \sin\theta'_W) Z^A_{i1} Z^H_{j1}$$
$$- (g_1 \cos\theta'_W \sin\theta_W + g_2 \cos\theta_W \cos\theta'_W - g_{YB} \sin\theta'_W) Z^A_{i2} Z^H_{j2} - 2 g_B \sin\theta'_W (Z^A_{i3} Z^H_{j3} - Z^A_{i4} Z^H_{j4})). \tag{D10}$$

$$g_{eeA1} = \frac{1}{\sqrt{2}} \sum_{b=1}^{3} \sum_{a=1}^{3} Y_{e,ab} Z^A_{k1},$$
$$g_{eeA2} = -\frac{1}{\sqrt{2}} \sum_{b=1}^{3} \sum_{a=1}^{3} Y^*_{e,ab} Z^A_{k1}. \tag{D11}$$

$$g_{Z'Ah} = \frac{1}{2}(-((g_1 \sin\theta_W + g_2 \cos\theta_W) \sin\theta'_W + g_{YB} \cos\theta'_W) Z^A_{i1} Z^H_{j1}$$
$$+ (g_1 \sin\theta'_W \sin\theta_W + g_2 \cos\theta_W \sin\theta'_W + g_{YB} \cos\theta'_W) Z^A_{i2} Z^H_{j2} - 2(g_B \cos\theta'_W (Z^A_{i3} Z^H_{j3} - Z^A_{i4} Z^H_{j4})). \tag{D12}$$

[1] S. Khalil and H. Okada, Phys. Rev. D **79**, 083510 (2009).
[2] A. Elsayed, S. Khalil, and S. Moretti, Phys. Lett. B **715**, 208 (2012).
[3] G. Brooijmans *et al.*, arXiv:1203.1488.
[4] L. Basso and F. Staub, Phys. Rev. D **87**, 015011 (2013).
[5] L. Basso, A. Belyaev, D. Chowdhury, M. Hirsch, S. Khalil, S. Moretti, B. O'Leary, W. Porod, and F. Staub, Comput. Phys. Commun. **184**, 698 (2013).
[6] A. Elsayed, S. Khalil, S. Moretti, and A. Moursy, Phys. Rev. D **87**, 053010 (2013).
[7] S. Khalil and S. Moretti, Rep. Prog. Phys. **80**, 036201 (2017).
[8] S. Khalil and A. Masiero, Phys. Lett. B **665**, 374 (2008).
[9] T. R. Dulaney, P. Fileviez Perez, and M. B. Wise, Phys. Rev. D **83**, 023520 (2011).
[10] V. Barger, P. Fileviez Perez, and S. Spinner, Phys. Rev. Lett. **102**, 181802 (2009).
[11] Lorenzo Basso, Adv. High Energy Phys. **2015**, 980687 (2015).
[12] A. Hammad, S. Khalil, and S. Moretti, Phys. Rev. D **93**, 115035 (2016).
[13] Tadashi Kon, Takuto Nagura, and Takahiro Ueda, Phys. Rev. D **99**, 095027 (2019).
[14] O. J. P. Eboli, G. C. Marques, S. F. Novaes, and A. A. Natale, Phys. Lett. B **197**, 269 (1987).
[15] E. W. N. Glover and J. J. van der Bij, Nucl. Phys. **B309**, 282 (1988); D. A. Dicus, C. Kao, and S. S. D. Willenbrock, Phys. Lett. B **203**, 457 (1988); T. Plehn, M. Spira, and P. M. Zerwas, Nucl. Phys. **B479**, 46 (1996); **B531**, 655(E) (1998).
[16] A. Djouadi, W. Kilian, M. Muhlleitner, and P. M. Zerwas, Eur. Phys. J. C **10**, 45 (1999).
[17] V. A. Iiyin, A. E. Pukhov, Y. Kurihara, Y. Shimizu, and T. Kaneko, Phys. Rev. D **54**, 6717 (1996).
[18] V. Barger and T. Han, Mod. Phys. Lett. A **05**, 667 (1990).
[19] A. Dobrovolskaya and V. Novikov, Z. Phys. C **52**, 427 (1991).
[20] D. A. Dicus, K. J. Kallianpur, and S. S. D. Willenbrock, Phys. Lett. B **200**, 187 (1988).
[21] A. Abbasabadi, W. W. Repko, D. A. Dicus, and R. Vega, Phys. Rev. D **38**, 2770 (1988).
[22] A. Gutirrez-Rodrguez, A. Hernndez-Ruz, and O. Sampayo, J. Phys. Soc. Jpn. **77**, 094101 (2008).
[23] E. Asakawa, D. Harada, and S. Kanemura, Phys. Rev. D **82**, 115002 (2010).
[24] A. Arhrib, R. Benbrik, and C.-W. Chiang, Phys. Rev. D **77**, 115013 (2008).
[25] N. Sonmez, J. High Energy Phys. 02 (2014) 042.
[26] J. Strube (ILC Physics and Detector Study Collaboration), Nucl. Part. Phys. Proc. **273–275**, 2463 (2016).
[27] B. OLeary, W. Porod, and F. Staub, J. High Energy Phys. 05 (2012) 042.
[28] W. Abdallah, S. Khalil, and S. Moretti, Phys. Rev. D **91**, 014001 (2015).
[29] L. Basso, Adv. High Energy Phys. **2015**, 1 (2015).
[30] S. Khalil and C. S. Un, Phys. Lett. B **763**, 164 (2016).
[31] F. Staub, arXiv:0806.0538.
[32] F. Staub, Comput. Phys. Commun. **181**, 1077 (2010).
[33] F. Staub, Comput. Phys. Commun. **182**, 808 (2011).
[34] F. Staub, Comput. Phys. Commun. **184**, 1792 (2013).
[35] F. Staub, Comput. Phys. Commun. **185**, 1773 (2014).
[36] Janusz Rosiek, arXiv:hep-ph/9511250.
[37] B. Holdom, Phys. Lett. **166B**, 196 (1986).
[38] T. Matsuoka and D. Suematsu, Prog. Theor. Phys. **76**, 901 (1986).
[39] F. del Aguila, G. D. Coughlan, and M. Quiros, Nucl. Phys. **B307**, 633 (1988).
[40] F. del Aguila, J. A. Gonzalez, and M. Quiros, Nucl. Phys. **B307**, 571 (1988).
[41] F. del Aguila, G. D. Coughlan, and M. Quiros, Nucl. Phys. **B312**, 751 (1989).
[42] R. Foot and X. G. He, Phys. Lett. B **267**, 509 (1991).

[43] K. S. Babu, C. F. Kolda, and J. March-Russell, Phys. Rev. D **57**, 6788 (1998).
[44] J. L. Yang, T. F. Feng, and H. B. Zhang, Phys. Rev. D **99**, 015002 (2019).
[45] J. L. Yang, T. F. Feng, S.-M. Zhao, R.-F. Zhu, X.-Y. Yang, and H.-B. Zhang, Eur. Phys. J. C **78**, 714 (2018).
[46] J. E. Camargo-Molina, B.O'Leary, and W. Porod, Phys. Rev. D **88**, 015033 (2013).
[47] O. A. Koval, L. R. Boyko, and N. Huseynov, AIP Conf. Proc. **2163**, 030008 (2019).
[48] N. Mebarki, M. Djouala, J. Mimouni, and H. Aissaoui, J. Phys. Conf. Ser. **1258**, 012011 (2019).
[49] A. Papaefstathiou, G. Tetlalmatzi-Xolocotzi, and M. Zaro, Eur. Phys. J. C **79**, 947 (2019).
[50] C. Partignani *et al.* (PDG Collaboration), Chin. Phys. C **40**, 100001 (2016); and 2017 update.
[51] J. L. Yang, T. F. Feng, H. B. Zhang, G. Z. Ning, and X. Y. Yang, Eur. Phys. J. C **78**, 438 (2018).
[52] ATLAS Collaboration, Report No. ATLAS-CONF-2016-045.
[53] G. Cacciapaglia, C. Csaki, G. Marandella, and A. Strumia, Phys. Rev. D **74**, 033011 (2006).
[54] M. Carena, A. Daleo, B. A. Dobrescu, and T. M. P. Tait, Phys. Rev. D **70**, 093009 (2004).
[55] ATLAS Collaboration, Phys. Rev. D **87**, 012008 (2013).
[56] CMS Collaboration, J. High Energy Phys. 10 (2012) 018.
[57] C. S. Un and O. Ozdal, Phys. Rev. D **93**, 055024 (2016).
[58] S. Kumar, P. Poulose, R.fiqul Rahaman, and R. K. Singh, Int. J. Mod. Phys. A **34**, 1950094 (2019).
[59] A. Djouadi, W. Kilian, M. Muhlleitner, and P. M. Zerwas, arXiv:hep-ph/0001169.